\documentclass[letterpaper,superscriptaddress,twocolumn,accepted=2024-01-01]{quantumarticle}

\pdfoutput=1

\usepackage{amsfonts}
\usepackage{amsmath}
\usepackage{amssymb}
\usepackage{graphicx}
\usepackage{mathtools}
\usepackage{hyperref}

\providecommand{\U}[1]{\protect\rule{.1in}{.1in}}
\newtheorem{theorem}{Theorem}

\newtheorem{lemma}[theorem]{Lemma}

\newtheorem{proposition}[theorem]{Proposition}

\newenvironment{proof}[1][Proof]{\noindent\textbf{#1.} }{\ \rule{0.5em}{0.5em}}

\allowdisplaybreaks

\newcommand{\bra}[1]{\langle{#1}|}
\newcommand{\ket}[1]{|{#1}\rangle}

\newcommand{\ketbra}[2]{|{#1}\rangle\!\langle{#2}|}

\newcommand{\beq}{\begin{equation}}
\newcommand{\beql}[1]{\begin{equation}\label{#1}}
\newcommand{\eeq}{\end{equation}}
\newcommand{\eeqp}{\, .\end{equation}}
\newcommand{\eeqc}{\, ,\end{equation}}

\newcommand*{\Tr}{\operatorname{Tr}}

\begin{document}
\title{Multivariate trace estimation in constant quantum depth}
\author{Yihui Quek}
\affiliation{Department of Mathematics, Massachusetts Institute of Technology, Cambridge MA 02139}
\affiliation{Dahlem Center for Complex Quantum Systems,
Freie Universit\"at Berlin, 14195 Berlin, Germany}
\affiliation{Information Systems Laboratory, Stanford University, Palo Alto, CA 94305, USA}
\author{Eneet Kaur}
\affiliation{Cisco Quantum Lab, Los Angeles, USA}
\affiliation{Institute for Quantum Computing and Department of Physics and Astronomy,
University of Waterloo, Waterloo, Ontario, Canada N2L 3G1}
\author{Mark M. Wilde}
\affiliation{School of Electrical and Computer Engineering, Cornell University, Ithaca, New York 14850, USA}
\affiliation{Hearne Institute for Theoretical Physics, Department of Physics and Astronomy,
and Center for Computation and Technology, Louisiana State University, Baton
Rouge, Louisiana 70803, USA}

\begin{abstract}
There is a folkloric belief that a depth-$\Theta(m)$ quantum circuit is needed to estimate the trace of the product of $m$ density matrices (i.e., a multivariate trace), a subroutine crucial to applications in condensed matter and quantum information science. We prove that this belief is overly conservative by constructing a constant quantum-depth  circuit for the task, inspired by the method of Shor error correction. Furthermore, our circuit demands only local gates in a two dimensional circuit -- we show how to implement it in a highly parallelized way on an architecture similar to that of Google's {\em Sycamore} processor. With these features, our algorithm brings the central task of multivariate trace estimation closer to the capabilities of near-term quantum processors. We instantiate the latter application with a theorem on estimating nonlinear functions of quantum states with ``well-behaved" polynomial approximations.
\end{abstract}
\date{}
\maketitle
\tableofcontents

\section{Introduction}

The task of estimating quantities like 
\beq\label{eq:pt}
\Tr[\rho_1\cdots \rho_m] \qquad \text{`Multivariate traces'}
\eeq
given access to copies of the quantum states $\rho_1$ through~$\rho_m$ is a fundamental building block in quantum information science. This subroutine, which we call `multivariate trace estimation', opens the door to estimating nonlinear functions of quantum states \cite{EAOHHK02,Brun04}, such as quantum distinguishability measures \cite{Buhrman_2001}, integer R\'enyi entropies \cite{johri17,EVDCZ18,VEDCZ18}, entanglement measures \cite{Ekert89,LLW04,infiniteauthors19} or testing sets of states for linear independence, coherence, and imaginarity \cite{OBG21}. This estimation procedure is a component of many quantum protocols such as quantum fingerprinting \cite{Buhrman_2001} and quantum digital signatures~\cite{gottesman_digital}. Let us note that multivariate traces are also known as {\em Bargmann invariants} \cite{OBG21,CW16,B64}.

When $\rho_i = \varrho$ for all $i \in [m]$ in \eqref{eq:pt}, an important application of multivariate trace estimation is to {\em entanglement spectroscopy} \cite{johri17} -- deducing the full set of eigenvalues $\{\lambda_1,\ldots,\lambda_D\}$ (the `spectrum') of $\varrho$, where~$\varrho$ is the reduced state of a bipartite pure state (that is, $\varrho = \Tr_{B}(\psi_{AB})$ with $\psi_{AB}$  a bipartite pure state). The spectrum unlocks a wealth of information about the properties of $\varrho$. The smallest eigenvalue of $\varrho$ diagnoses whether $\psi$ is separable or entangled \cite{Ekert89}. In addition, the {\em inverse} of the smallest eigenvalue acts as a condition number for many quantum algorithms that manipulate quantum states (see for instance \cite{HHL,gilyen2018QSingValTransf,Petz}), and it constrains their runtime. 
The entanglement spectrum is also useful to identify topological order \cite{Masaki2010,Yao2010,Lukasz2010,Hui2008}, emergent irreversibility~\cite{Claudio2014},  quantum phase transitions~\cite{Chiara2012}, and to determine if the system obeys an area law~\cite{Eisert2010}.

With the wealth of applications described above, there has been much interest in bringing multivariate trace estimation within the reach of near-term quantum hardware. 
A glimmer of hope in this regard is the observation that quantities like that in \eqref{eq:pt} can be estimated {\em without} the need for full-state tomography. In the quantum information sphere, one of the progenitors of this line of thinking was Ref.~\cite{EAOHHK02}, which proposed a method leveraging the following well-known identity (related to the replica trick originating in spin glass theory \cite{MPV86}):
\beq\label{eq:id}
\Tr[W^{\pi} (\rho_1 \otimes \cdots \otimes \rho_m)] = \Tr[\rho_1\cdots \rho_m]
\eeqc
where the right-hand-side is the multivariate trace we would like to estimate, and $W^{\pi}$ is a unitary representation of the cyclic shift permutation 
\beq
\pi \coloneqq  (1, 2, \ldots, m)
\eeqp
Here, $\pi$ represents a cyclic permutation that sends 1 to 2, 2 to 3, and so forth. 
That is to say, multivariate trace estimation can be accomplished by estimating the real and imaginary parts of the {\em cyclic shift} operator in \eqref{eq:id}, using quantum hardware. This identity subsequently became the backbone of many proposals \cite{Ekert89, johri17,Yirka2021qubitefficient, Suba__2019,OBG21} for multivariate trace estimation with yet more near-term constraints. However, there appears to be a lack of clarity regarding the actual resource requirements of these methods. Refs.~\cite{Yirka2021qubitefficient, Suba__2019, johri17, Ekert89,OBG21} have all suggested that a quantum circuit whose depth is linear in $m$ is needed to perform the task. Ref.~\cite{OBG21} has additionally proposed a logarithmic-depth circuit for this task, but it requires all-to-all qubit connectivity in a quantum computer, which is unlikely to be available in the near-term.

In this paper, we show that these characterizations are overly-conservative: multivariate trace estimation can be implemented in {\em constant} quantum depth -- the shortest depth one could hope for -- while simultaneously respecting near-term quantum architecture constraints. In particular, we require only two-dimensional nearest-neighbor connectivity (already a feature of Google's {\em Sycamore} processor \cite{Arute2019}), linearly-many controlled two-qubit gates and a linear amount of classical pre-processing. We invoke ideas from Shor error correction \cite{S96} to construct a circuit that achieves this claim (Section~\ref{sec:ourconstruction}), show how this circuit can be implemented in a highly parallelized way on a two-dimensional~architecture similar to Google's (Section~\ref{sec:2D}), prove Theorem \ref{thm:estimate-trace-poly-func} about the statistical guarantees of the resulting estimator (Section~\ref{sec:guarantees}), and show that our method finds further application in estimating traces of `well-behaved' polynomial functions of density matrices (Section~\ref{sec:app}). The next section begins the technical part of our paper by reviewing the basic principle behind multivariate trace estimation.


\section{The principle: Using cyclic shifts for multivariate trace estimation}\label{sec:principle}

The principle behind our circuit construction is simple. To explain it, let us first recall a well known circuit construction \cite{EAOHHK02} for multivariate trace estimation  that has no clear realization on near-term quantum computers. The idea is to estimate the quantities $\operatorname{Re}
[\operatorname{Tr}[\rho_1\cdots \rho_m]]$ and $\operatorname{Im}
[\operatorname{Tr}[\rho_1\cdots \rho_m]]$ separately. 

The circuits to estimate both quantities are similar, and we now describe them. To estimate the real part $\operatorname{Re}
[\operatorname{Tr}[\rho_1\cdots \rho_m]]$, we proceed according to the following steps: 
\begin{enumerate}

\item Prepare a qubit in the $|+\rangle
\coloneqq(|0\rangle+|1\rangle)/\sqrt{2}$ state and adjoin to it the state $\rho_1\otimes \cdots \otimes \rho_m$.

\item Perform a controlled cyclic permutation
unitary gate, defined as%
\begin{equation}
|0\rangle\!\langle0|\otimes I^{\otimes m}+|1\rangle\!\langle1|\otimes
W^{\pi}.
\end{equation}

\item Measure the first qubit in the basis $\{|+\rangle,|-\rangle\}$, where
$|-\rangle\coloneqq(|0\rangle-|1\rangle)/\sqrt{2}$, and record the outcome
$X=+1$ if the first outcome $|+\rangle$ is observed and $X=-1$ if the second
outcome $|-\rangle$ is observed. 

\item Repeat Steps 1 to 3 a number of times equal to $N\coloneqq O(\varepsilon^{-2} \log\delta^{-1}) $ and return $\hat{X} \coloneqq \frac{1}{N}\sum_{i=1}^N X_i$, where $X_i$ is the outcome of the $i$-th repetition of Step 3.  
\end{enumerate} 

It is known that
\beq\label{eq:av}
\mathbb{E}[X] = \operatorname{Re}
[\operatorname{Tr}[\rho_1\cdots \rho_m]]
\eeqc
and thus $\hat{X}$ computed in Step 4 is an empirical estimate of the desired quantity. That is, by invoking the well known Hoeffding inequality, it is guaranteed, for $\varepsilon > 0$ and $\delta \in (0,1)$, that
\begin{equation}
\Pr(|\hat{X} - \operatorname{Re}[\operatorname{Tr}[\rho_1\cdots \rho_m]]| \leq \varepsilon) \geq 1- \delta .
\end{equation}
For completeness, we recall the Hoeffding inequality now:

\begin{lemma}
[Hoeffding \cite{H63}]\label{lem:hoeffding}
Suppose that we are given $n$~independent
samples $Y_{1}$, \ldots, $Y_{n}$ of a bounded random variable~$Y$ taking
values in $\left[  a,b\right]  $ and having  mean~$\mu$. Set%
\begin{equation}
\overline{Y^{n}}\coloneqq \frac{1}{n}\left(  Y_{1}+\cdots+Y_{n}\right)
\end{equation}
to be the sample mean. Let $\varepsilon > 0  $ be the desired
accuracy, and let $1-\delta$ be the desired success probability, where
$\delta\in\left(  0,1\right)  $. Then
\begin{equation}
\Pr\!\left(  \left\vert \overline{Y^{n}}-\mu\right\vert \leq\varepsilon
\right)  \geq1-\delta,
\end{equation}
if
\begin{equation}
n\geq\frac{M^{2}}{2\varepsilon^{2}}\ln\!\left(  \frac{2}{\delta}\right)  ,
\end{equation}
where $M\coloneqq b-a$.
\end{lemma}

To see that \eqref{eq:av} holds, note that in the special case when all the states are pure, i.e., $\rho_i = \ketbra{\psi_i}{\psi_i}$, the input to the circuit is an $m$-partite pure state $\ket{\psi^{(m)}} \coloneqq \ket{\psi_1}\otimes \cdots \otimes \ket{\psi_m}  $, and so
\begin{align}
&\Pr(X=+1) \nonumber \\
&= \left\|(\bra{+} \otimes I) \frac{1}{\sqrt{2}} \left(  |0\rangle|\psi^{(m)}\rangle
+|1\rangle W^{\pi}|\psi^{(m)}\rangle\right)\right\|_2^2\\
&= \frac{1}{4} \left\Vert |\psi^{(m)}\rangle + W^{\pi}|\psi^{(m)}\rangle \right\Vert_2^2\\
&= \frac{1}{4}\left( 2+ \bra{\psi^{(m)}}W^{\pi}|\psi^{(m)}\rangle + \bra{\psi^{(m)}}{(W^{\pi})}^{\dagger}|\psi^{(m)}\rangle\right)\\
&= \frac{1}{2}(1+ \operatorname{Re}[\Tr[W^{\pi} \ketbra{\psi^{(m)}}{\psi^{(m)}} ]]) \\
&= \frac{1}{2}(1 + \operatorname{Re}[\Tr[\rho_1\cdots \rho_m]])\label{eq:bias},
\end{align}
where in the last equality we have used the well-known identity in \eqref{eq:id}. Similarly, we have that
\begin{equation}
\Pr(X=-1) = \frac{1}{2}(1 - \operatorname{Re}[\Tr[\rho_1\cdots \rho_m]]),
\end{equation}
so that
\begin{align}
\mathbb{E}[X] & = (+1)\Pr(X=+1) + (-1) \Pr(X=-1)\\
& = \operatorname{Re}[\Tr[\rho_1\cdots \rho_m]].
\end{align}
Eq.~\eqref{eq:av}, which asserts that the conclusion in \eqref{eq:bias} still holds when each $\rho_i$ is a mixed state, follows by convexity (i.e., that every mixed state can be written as a convex combination of pure states).

To estimate the second quantity
$\operatorname{Im}[\operatorname{Tr}[\rho_1\cdots \rho_m]]$, a simple variation of the above argument suffices. The technique is identical, except that the final measurement is in the basis $\{|+_{Y}\rangle,|-_{Y}\rangle\}$, where
$|\pm_{Y}\rangle\coloneqq\left(|0\rangle\pm i|1\rangle\right)/\sqrt{2}$. 

\section{Our circuit construction}
\label{sec:ourconstruction}

We propose a variation of the above method, which instead estimates $\operatorname{Tr}[\rho_1\cdots \rho_m]$ in {\em constant} quantum depth.  This makes the circuit more amenable to run on near-term quantum processors. This circuit is depicted in Figure~\ref{fig:cyclic-shift-test} for $m=8$ --- it has some
similarities with the method of Shor error correction from fault-tolerant
quantum computation \cite{S96} (see also Figure~2 of \cite{G10}).

The key enabler of our depth reduction is the replacement of the $\ket{+} = \frac{1}{\sqrt{2}} (\ket{0} + \ket{1})$ state at the single qubit control wire in the circuit in \cite{EAOHHK02}, with an $\lfloor m/2 \rfloor$-party
GHZ\ state
\begin{equation}\label{eq:GHZ}
|\Phi_{\text{GHZ}}^{\lfloor m/2 \rfloor}\rangle\coloneqq \frac{1}{\sqrt{2}}\left(  |0\rangle^{\otimes
\lfloor m/2 \rfloor}+|1\rangle^{\otimes \lfloor m/2 \rfloor}\right)  
\end{equation}
on $\lfloor m/2 \rfloor$ control wires. This modification was also proposed in \cite{OBG21}, but their circuit for it is of logarithmic depth. In any case, this modification allows the number of controls to the permutation to increase to $\lfloor m/2 \rfloor$, which is half the input size. In turn, it paves the way for an implementation of multivariate trace estimation in quantum depth two using parallelized controlled-SWAP gates. 

In order to achieve an overall constant quantum depth, in Section~\ref{subsec:GHZ-prep-constant-depth} we show that constant quantum depth suffices to generate the input GHZ state in \eqref{eq:GHZ}. Then, in Section~\ref{subsec:permutation-constant-depth}, we show how to implement the permutation $W^{\pi}$, again in constant quantum depth. Finally, in Section~\ref{subsec:explicit}, we describe our full estimator and the accompanying circuit. 

\subsection{Preparing GHZ\ states in constant quantum depth} \label{subsec:GHZ-prep-constant-depth}

We now describe two methods to generate the GHZ state in constant quantum depth. The first method has been discussed previously in \cite{WKST19} and is a constant quantum depth
circuit assisted by measurements, classical feedback, and a logarithmic depth
classical circuit. See also the discussion after \cite[Theorem~1.1]{LG21}. The second is a variation of the first, which additionally allows for qubit resets to make more efficient usage of qubits.

\textbf{Method 1.} We begin by discussing the first approach, which generates an
$r$-party GHZ state using a constant quantum depth circuit assisted by
measurements, classical feedback, and a logarithmic depth classical circuit.
We proceed according to the following steps:

\begin{enumerate}
\item Generate $r-1$ Bell states; i.e., each pair is in the state%
\begin{equation}
|\Phi^{+}\rangle\coloneqq \frac{1}{\sqrt{2}}\left(  |00\rangle+|11\rangle\right)  .
\end{equation}

\item Perform controlled-NOT gates between the second qubit in each Bell state and the first qubit of the following one. 

That is, apply a controlled-NOT from qubit $2k$ to qubit $2k+1$ for all $k\in \{1, \ldots, r-2\}$.

\item Measure the target of each controlled-NOT gate (all odd-numbered qubits except the first qubit) in the computational basis.

\item Controlled on the measurement outcome $b_1,\ldots, b_{r-2}$, apply a tensor product of Pauli~$X$ operators as a correction to all even-numbered qubits except the second qubit. 

That is, apply $X^{b_1\oplus\cdots \oplus b_{k-1}}$ to qubit $2k$ for $k \in \{ 2, \ldots, r-1\}$. 
\end{enumerate}

\noindent This procedure prepares an $r$-party GHZ state on qubits $1, 2, 4, 6, \ldots, 2(r-1)$.

We now show in detail that the scheme prepares an $r$-party GHZ state. Consider that the
initial state can be written as%
\begin{align}
&  \bigotimes\limits_{i=1}^{r-1}|\Phi^{+}\rangle\nonumber\\
&  =\frac{1}{\sqrt{2^{r-1}}}\bigotimes\limits_{i=1}^{r-1}\sum_{x_{i}%
\in\left\{  0,1\right\}  }|x_{i},x_{i}\rangle\\
&  =\frac{1}{\sqrt{2^{r-1}}}\sum_{x_{1},\ldots,x_{r-1}\in\left\{  0,1\right\}
}|x_{1},x_{1}\rangle\otimes\bigotimes\limits_{i=2}^{r-1}|x_{i},x_{i}\rangle.
\end{align}
After step~2, the state becomes%
\begin{equation}
\frac{1}{\sqrt{2^{r-1}}}\sum_{x_{1},\ldots,x_{r-1}\in\left\{  0,1\right\}
}|x_{1},x_{1}\rangle\otimes\bigotimes\limits_{i=2}^{r-1}|x_{i}\oplus
x_{i-1},x_{i}\rangle.
\end{equation}
After step~3, the $(r-2)$-bit string $b_{1}\cdots b_{r-2}$ is obtained from
the measurements, where%
\begin{align}
b_{1}  &  =x_{2}\oplus x_{1},\\
b_{2}  &  =x_{3}\oplus x_{2},\\
&  \cdots,\\
b_{r-2}  &  =x_{r-1}\oplus x_{r-2},
\end{align}
and this projects the state onto the following state:
\begin{align}
&  \frac{1}{\sqrt{2}}\sum_{x_{1}\in\left\{  0,1\right\}  }|x_{1},x_{1}%
\rangle \otimes |b_1, x_2\rangle\otimes|b_2, x_{3}\rangle\cdots\otimes |b_{r-2}, x_{r-1}\rangle\nonumber\\
&= \frac{1}{\sqrt{2}}\sum_{x_{1}\in\left\{  0,1\right\}  }|x_{1},x_{1}%
\rangle \otimes |b_1, b_1 \oplus x_1\rangle\otimes\nonumber\\
&  \qquad|b_2, b_{1}\oplus b_{2}\oplus x_{1}\rangle\cdots \otimes |b_{r-2}, b_{1}\oplus\cdots\oplus
b_{r-2}\oplus x_{1}\rangle\\
&  =(  I\otimes I\otimes I \otimes  X^{b_{1}}\otimes I \otimes X^{b_{1}\oplus b_{2}}%
\otimes\cdots \nonumber \\
& \qquad\qquad\qquad\qquad \otimes I \otimes X^{b_{1}\oplus\cdots\oplus b_{r-2}})
\times\nonumber\\
&  \qquad\left(  \frac{1}{\sqrt{2}}\sum_{x_{1}\in\left\{  0,1\right\}  }%
|x_{1},x_{1}\rangle|b_1, x_{1}\rangle|b_2, x_{1}\rangle\cdots|b_{r-2}, x_{1}\rangle\right) .
\end{align}
The $r$-party GHZ\ state on qubits $1, 2, 4,6,\ldots , 2(r-1)$ is then recovered by performing the following
correction operation:%
\begin{equation}
I\otimes I\otimes I \otimes  X^{b_{1}}\otimes I \otimes X^{b_{1}\oplus b_{2}}%
\otimes\cdots I \otimes X^{b_{1}\oplus\cdots\oplus b_{r-2}}.
\label{eq:ghz-prep-correction}
\end{equation}

\begin{figure}
\begin{center}
\includegraphics[
width=\columnwidth
]{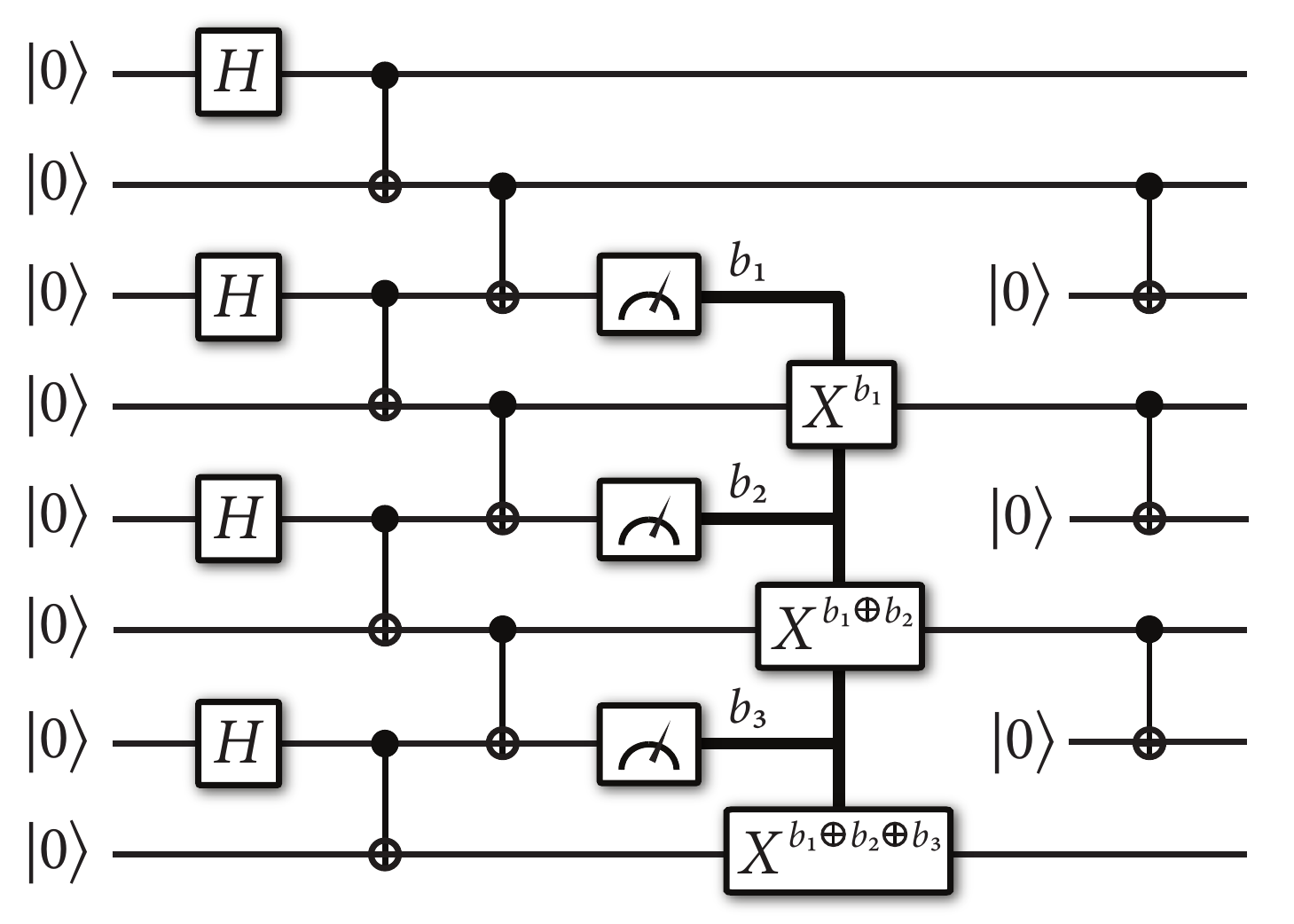}
\end{center}
\caption{A constant-depth quantum circuit for preparing an eight-party GHZ state, assisted by
measurement, classical feedback, and qubit resets. Method 1 consists of all the steps depicted, except for the qubit resets and final controlled-NOTs (but only prepares a five-party state). In Method 2, the measured qubits are additionally reset to the $|0\rangle$ state and connected by controlled-NOTs so that a larger, eight-party state can be prepared instead.}%
\label{fig:ghz-prep-qubit-resets}%
\end{figure}

The leftmost part of Figure \ref{fig:ghz-prep-qubit-resets} (all except the qubit resets and final controlled-NOTs) depicts this procedure.

\textbf{Method 2.} The second method is very similar to the one just described. The
only difference is that we additionally perform qubit resets on the measured qubits and
then controlled-NOTs from the nearest neighbor qubits in the GHZ\ state to the
reset qubits. This scheme is depicted in Figure~\ref{fig:ghz-prep-qubit-resets}. It prepares a GHZ state on a number of parties equal to the number of input qubits (hence a $2(r-1)$-party GHZ state, as opposed to the $r$-party GHZ state without the qubit resets). 

\subsection{Multiply-controlled cyclic permutation in constant quantum depth} 

\label{subsec:permutation-constant-depth}

Rather than implement a controlled-$W^{\pi}$ gate, we instead implement a multiply-controlled-$W^{\pi}$ gate, in order to reduce the depth of this part of the circuit from linear to constant.  Our implementation of the multiply-controlled-$W^{\pi}$ gate in constant depth is based on the observation that there is a particularly convenient way to decompose a cyclic shift into a product of transpositions, as shown in  \cite[Eqs.~(4.2)--(4.3)]{GB00}. We state this observation here, along with a brief proof, for completeness:
\begin{proposition}The following decomposition holds
\label{prop:decomposition}
\begin{align}
&(1, \ldots, m) = \nonumber\\
&\left\{\begin{array}{ll}
      \displaystyle \prod_{l=2}^{m/2} (l, m+2-l) \prod_{k=1}^{m/2} (k , m+1-k)  & : m \text{ even} \\
      \displaystyle \prod_{l=2}^{\lceil m/2 \rceil} (l, m+2-l) \prod_{k=1}^{\lfloor m/2 \rfloor} (k , m+1-k) & : m \text{ odd}\label{eq:decomposition}\\
\end{array} \right.
\end{align}
where all arithmetic is modulo $m$.
\end{proposition}
So, for instance, when $m=8$ (the case in Figure \ref{fig:cyclic-shift-test}), Eq.~\eqref{eq:decomposition} reduces to
\beq
(1,\ldots,8) = (2,8) (3,7) (4,6) \quad (1,8) (2,7) (3,6) (4,5)
\eeqp
\begin{proof}
For $m$ even, the first transposition sends $k$ to $m+1-k$ for every $k\in [m]$. If $k = m$, it gets sent to $1$ by the first transposition and is not acted on by the second transposition. Otherwise, the second transposition sends $m+1-k$ to $m+2-(m+1-k) = k+1$, so the overall effect is to send $k \rightarrow k+1$ for all $k \in [m]$, as desired.

For $m$ odd, there are two indices that are involved in only one transposition: $k = m$, which, as before, gets sent to $1$ by the first transposition and is not acted on by the second transposition; and $k= \lceil m/2 \rceil$, which is transposed only in the second transposition, where it gets sent to $m+2-\lceil m/2 \rceil = \lceil m/2 \rceil +1$. All other indices are involved in the same two transpositions as described above. Thus, the overall effect is also $k \rightarrow k+1$ for all $k \in [m]$. 
\end{proof}

\medskip

Eq.~\eqref{eq:decomposition} says that, for a fixed $m$, the $m$-wise cyclic shift permutation can be decomposed into a product of two terms. Each term is itself a product of disjoint transpositions and every index gets transposed once per term.

This has a clear interpretation in terms of how to construct a quantum circuit to implement a multiply-controlled-$W^{\pi}$. Transposing two qubit labels can be achieved by applying a SWAP gate to the relevant qubits. Disjoint transpositions can thus be accomplished by implementing SWAP gates in parallel, in a single time step. Proposition \ref{prop:decomposition} thus implies that, for every $m$, the multiply-controlled-$W^{\pi}$ can be implemented in two time steps  (depth two), each of which performs $\lfloor m/2 \rfloor$ controlled SWAPs in parallel. 

We give a more explicit description of the circuit in the next subsection. Note that, as depicted in Figure~\ref{fig:cyclic-shift-test}, we have made another optimization for near-term feasibility: we adjoin the input states to the cyclic shift in a specific order that ensures that only nearest-neighbor states need to be swapped.

\subsection{Explicit circuit description}

\label{subsec:explicit}

\begin{figure}[ptb]
\begin{center}
\includegraphics[
width=\columnwidth
]{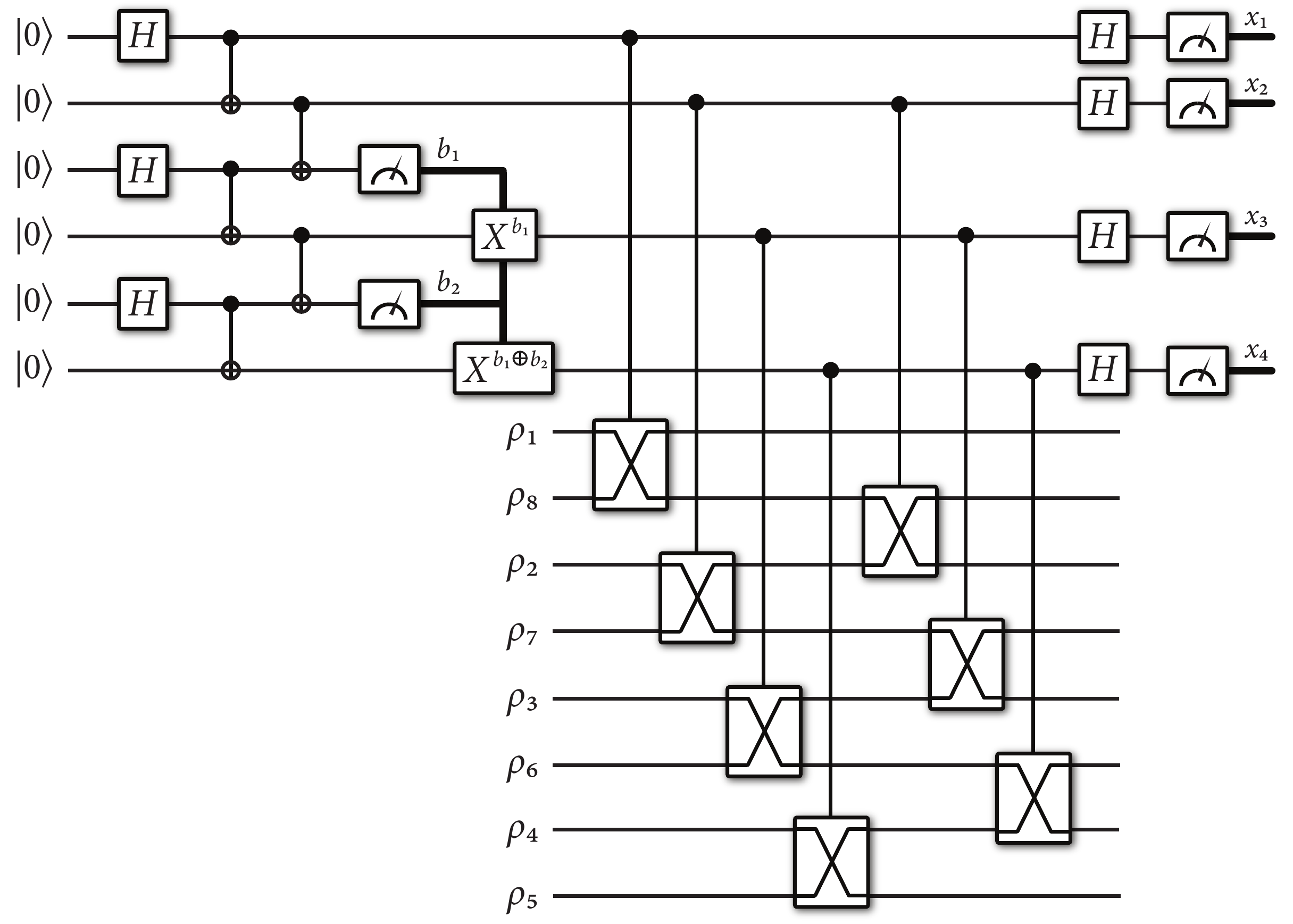}
\end{center}
\caption{The leftmost part of the circuit prepares a four-party GHZ\ state.
The middle part of the circuit performs a controlled cyclic-shift. The final
part of the circuit results in the classical bits $x_{1}$, $x_{2}$, $x_{3}$,
$x_{4}$, which are used to generate $r=\left(  -1\right)  ^{x_{1}+x_{2}%
+x_{3}+x_{4}}$. As argued in
Section~\ref{sec:guarantees}, the expectation of $r$
is equal to $\operatorname{Re}[\operatorname{Tr}[\rho_{1}\cdots\rho_{8}]]$, so
that this latter quantity can be estimated through repetition.}%
\label{fig:cyclic-shift-test}%
\end{figure}

We now put together the findings of the previous two subsections and describe our proposed technique for estimating $\Tr[\rho_1\cdots \rho_m]$ in the case that each local dimension
$d=2$ (i.e., each $\rho_i$ is a single-qubit state). The corresponding circuit is depicted in Figure~\ref{fig:cyclic-shift-test}. After that, we discuss how to generalize the construction to the case in which $d$ is a power of two, so that each local system consists of multiple qubits. The estimator works as follows:

\begin{enumerate}

\item Prepare an $\lfloor m/2 \rfloor$-party GHZ\ state using one of the constant quantum-depth circuit constructions described in the previous section. Let us call the $\lfloor m/2 \rfloor$ qubits of the GHZ\ state the control qubits and the $m$
states $\rho_1\otimes \cdots \otimes \rho_m$ the target qubits. 

\item To implement the multiply-controlled cyclic shift, 
 \begin{itemize}
 \item If $m$ is \textbf{even}, adjoin $\rho_{1}, \ldots, \rho_{m}$ to the GHZ\ state, in the order%
\begin{multline}
\rho_{1}\otimes\rho_{m}\otimes\rho_{2}\otimes\rho_{m-1}\otimes\rho
_{3}\otimes\cdots\\\otimes\rho_{m/2+2}\otimes\rho_{m/2}\otimes\rho_{m/2+1}.
\label{eq:even-number-states}
\end{multline}
Perform a
controlled-SWAP\ gate from the $i$th control qubit to target qubits $2i-1$ and $2i$, for all $i\in\left\{
1,\ldots,m/2\right\}  $. Now perform a controlled-SWAP\ from the $i$th control
qubit to target qubits $2i$ and $2i+1$ for $i\in\left\{  1,\ldots,m/2-1\right\}
$. 

\item If $m$ is \textbf{odd}, adjoin $\rho_{1}, \ldots, \rho_{m}$ to the GHZ\ state, in the order%
\begin{multline}
\rho_{1}\otimes\rho_{m}\otimes\rho_{2} \otimes \rho_{m-1}\otimes\rho_3 \otimes \\\cdots\otimes\rho_{\lceil m/2 \rceil-1}\otimes\rho_{\lceil m/2 \rceil+1}\otimes\rho_{\lceil m/2 \rceil}.
\label{eq:odd-number-states}%
\end{multline} 
Perform a controlled-SWAP\ gate from the $i$th control qubit to target qubits $2i-1$ and $2i$, for all $i\in\left\{
1,\ldots,\lfloor m/2 \rfloor\right\}  $. Now perform a controlled-SWAP\ from the $i$th control
qubit to target qubits $2i$ and $2i+1$ for $i\in\left\{  1,\ldots,\lfloor m/2 \rfloor\right\}
$. 
\end{itemize}
It can be checked that this prescription implements precisely \eqref{eq:decomposition}, with the states adjoined in a specific order (see \eqref{eq:even-number-states} or \eqref{eq:odd-number-states}) that allows only nearest neighbors to be swapped.
\item Perform a Hadamard on all $\lfloor m/2 \rfloor$ control qubits and measure them in
the computational basis, receiving outcomes $0$ or $1$. This has the effect of performing a measurement in the $X$ basis on each control qubit.  Let $X_{i}\in\left\{
0,1\right\}  $ denote the result of the $i$th measurement, for $i\in\left\{
1,\ldots,\lfloor m/2 \rfloor\right\}  $. Set $R=\left(  -1\right)  ^{\sum_{i=1}^{\lfloor m/2 \rfloor}X_{i}}$.

\item Repeat Steps 1 to 3 a number of times equal to $N\coloneqq  O(\varepsilon^{-2} \log \delta^{-1}) $. Compute $\hat{R} \coloneqq \frac{1}{N}\sum_{j=1}^N R_j$, where $R_j$ is the output of Step 3 on the $j$-th application of the circuit. $\hat{R}$ is our estimate for $\operatorname{Re}[\operatorname{Tr}[\rho_{1}\cdots\rho_{m}]]$.

\item To estimate $\operatorname{Im}[\operatorname{Tr}[\rho_{1}\cdots\rho_{m}]]$, repeat Steps 1 to 4, except that in Step 3, replace each Hadamard with $HS^{\dagger}$, where $S$ is the phase gate 
\begin{equation}
S \coloneqq 
  \begin{bmatrix}
    1 & 0 \\
    0 & i
  \end{bmatrix},
  \end{equation}
  before the measurement in the computational basis. This has the effect of performing a measurement in the $Y$ basis on each control qubit. Let $Y_{i}^{(j)} \in \{0, 1\}$ be the outcome of the measurement on the $i$-th qubit on the $j$-th application of the circuit. Set $J_j=\left(  -1\right)  ^{\sum_{i=1}^{\lfloor m/2 \rfloor}Y_{i}^{(j)}}$. Compute $\hat{J} = \frac{1}{N}\sum_{j=1}^N J_j$, which  is our estimate for  $\operatorname{Im}[\operatorname{Tr}[\rho_{1}\cdots\rho_{m}]]$.
  \item Output $\hat{T} = \hat{R} + i\hat{J}$. 
\end{enumerate}

Our proposed architecture is highly flexible and can be tailored to the availability of resources such as long coherence times and multi-qubit gates. This is evident in two ways: 

Firstly, we can smoothly trade off circuit width for the availability of entangling gates. At one extreme, observe that we could have implemented our circuit with only one control qubit (instead of $\lfloor m/2 \rfloor$) if we had at our disposal a highly-entangling gate: a single-qubit controlled simultaneous SWAP gate that swapped $\lfloor m/2 \rfloor$ pairs of qubits simultaneously. 
Such a gate would not be feasible in the near term, as it requires highly nonlocal interactions to implement in a real physical architecture. However, even gates that entangle only a {\em subset} of qubits afford us savings in circuit width: every additional controlled $k$-wise SWAP gate at our disposal allows for a reduction of circuit width by $k$, as $k$ fewer control qubits are necessary (hence the GHZ state needs to be on $k$ fewer parties). 

Secondly, to generalize the estimation of $\operatorname{Tr}[\rho
_{1}\rho_{2}\cdots\rho_{m}]$ beyond single-qubit states, we can either
increase the width or the depth of the circuit described above. Suppose that
$\rho_{1}, \ldots, \rho_{m}$ each consist of $p$ qubits. 

\begin{itemize}
\item We can increase
the width of the circuit by preparing GHZ\ states with $mp$ qubits and then
group these into $p$ groups of $m$ qubits each. Correspondingly, group the
$2m$ states $\rho_{1}, \ldots, \rho_{2m}$ into $p$ groups of $2m$ qubits,
where the $k$th group has the $k$th qubit of each state, for $k\in\left\{
1,\ldots,p\right\}  $. Then we perform controlled SWAPs as detailed above, for
each group. Finally, perform Hadamards on all of the $mp$ control qubits and
measure each of them in the computational basis. 
\item To increase the depth of the circuit, prepare an $m$-party GHZ\ state, and then
sequentially perform the controlled-SWAP tests for the $p$ groups of qubits,
so that the depth of the circuit increases by a factor of $p$. Then measure
the control qubits as before. 
\end{itemize}

\section{Implementing multivariate trace estimation on a two-dimensional architecture} \label{sec:2D}

\begin{figure*}
\begin{center}
\includegraphics[
width=1.52in
]{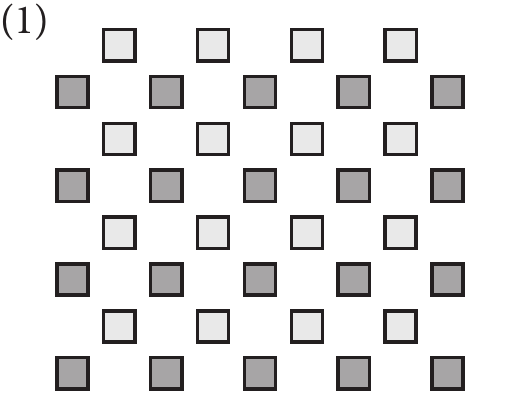}\hfil
\includegraphics[
width=1.52in
]{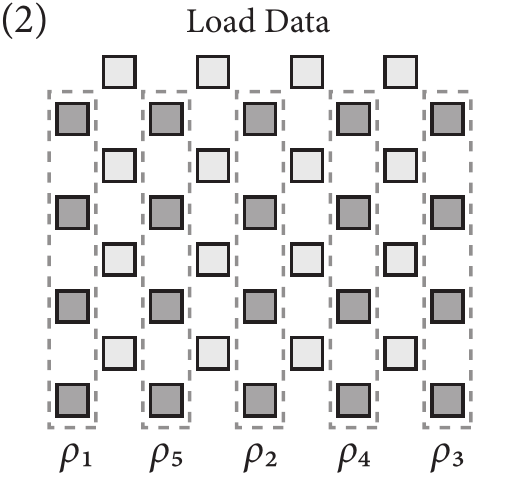}\hfil
\includegraphics[
width=1.52in
]{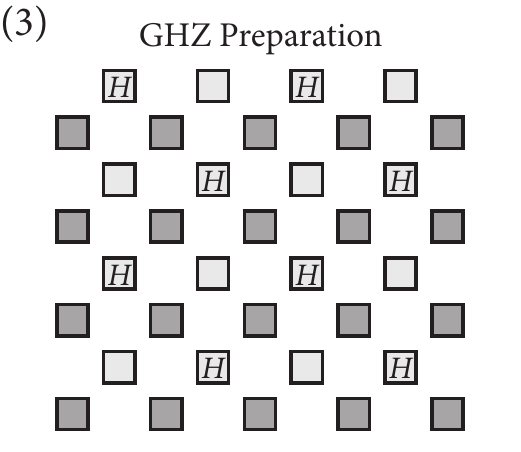}
\hfil
\includegraphics[
width=1.52in
]{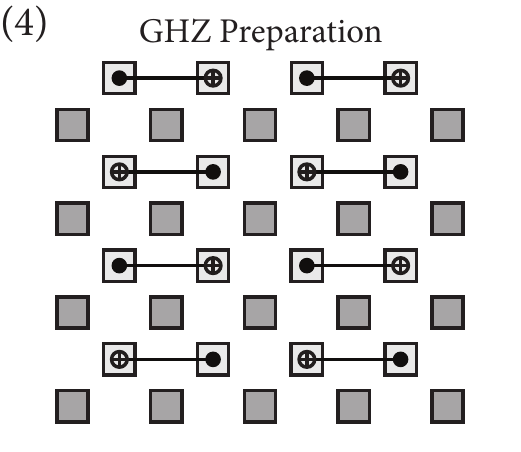}
\par\medskip
\includegraphics[
width=1.52in
]{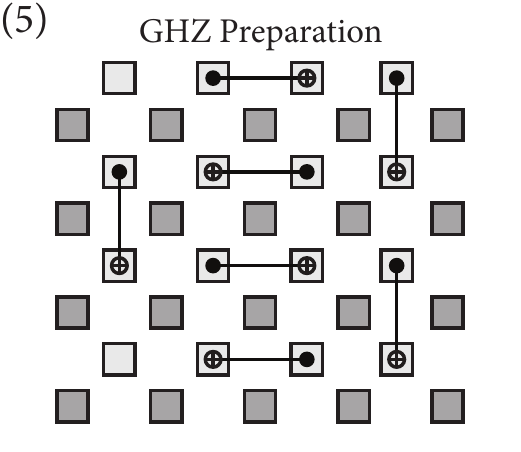}\hfil
\includegraphics[
width=1.72in
]{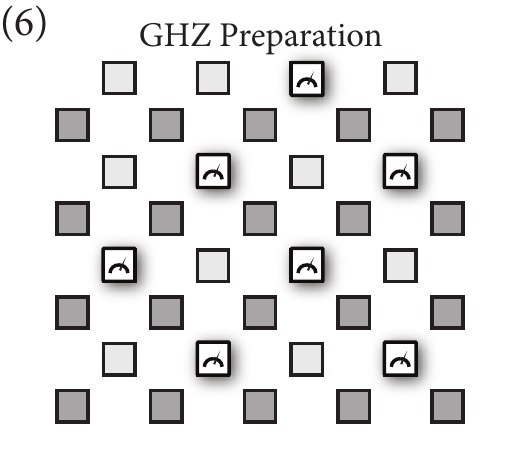}
\hfil
\includegraphics[
width=1.52in
]{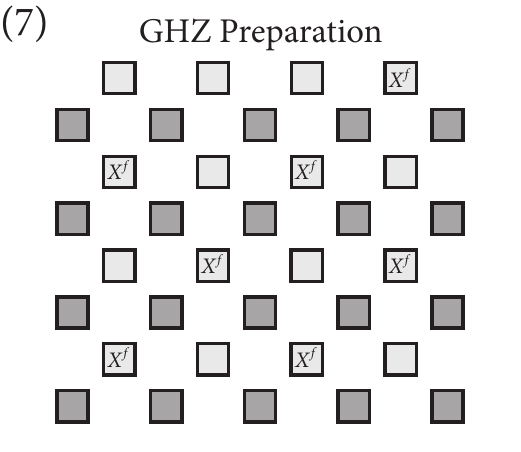}\hfil
\includegraphics[
width=1.52in
]{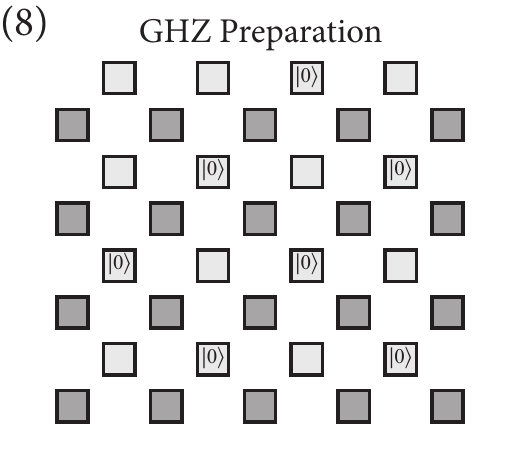}\par\medskip
\includegraphics[
width=1.72in
]{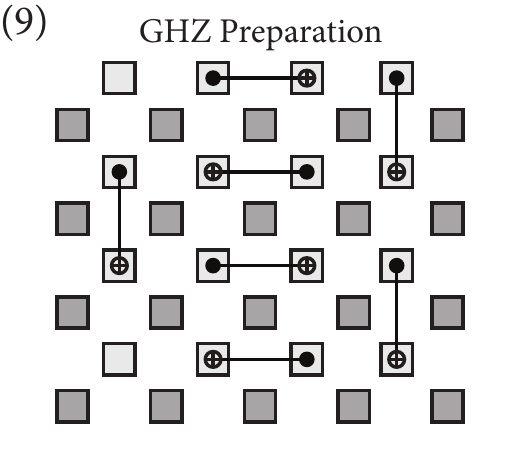}
\hfil
\includegraphics[
width=1.52in
]{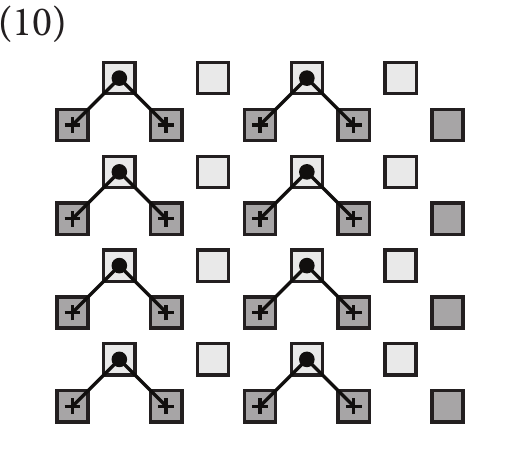}
\hfil 
\includegraphics[
width=1.52in
]{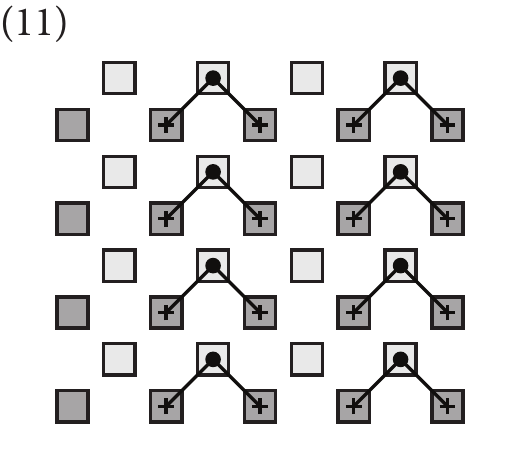}
\hfil
\includegraphics[
width=1.52in
]{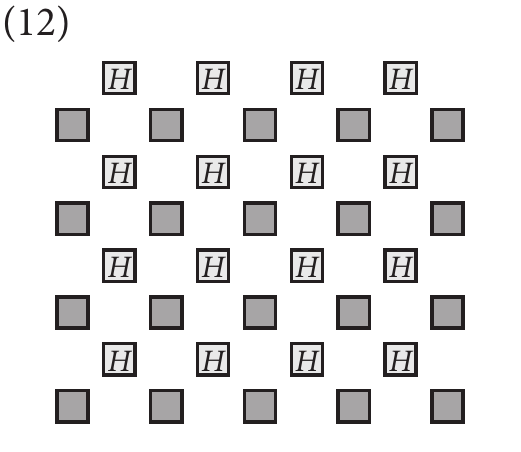}
\par\medskip
\includegraphics[
width=1.52in
]{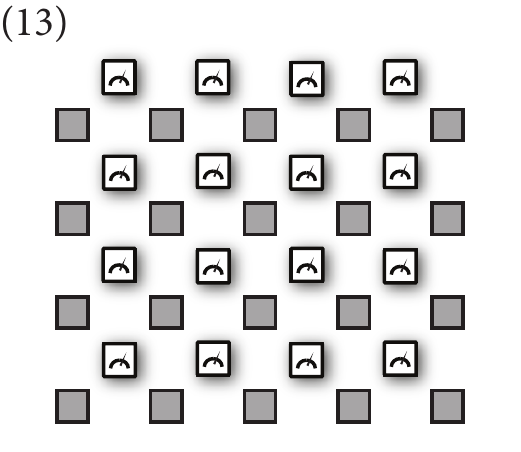}
\end{center}
\caption{(1) The squares in light grey represent control qubits, and the squares in dark grey represent data qubits. In this example, there are five data states involved, each consisting of four qubits. (2) The quantum data is loaded during this stage, which we note here can be conducted in parallel with the preparation of the GHZ state in the control qubits. The state $\rho_i$, for $i \in \{1, \ldots, 5\}$, is a four-qubit state that occupies the indicated column  of the data qubits in dark grey. The light grey control qubits are prepared in the all zeros state. (3) First step of the preparation of the GHZ state of the control qubits. Every other control qubit has a Hadamard gate applied in parallel. (4) Every pair of control qubits has CNOT gates applied in parallel. (5) Every other pair of control qubits has CNOT gates applied in parallel. (6) Starting from the third control qubit from the top left, every other control qubit is measured, and the measurement outcome is stored in a binary vector $b_1, \ldots, b_7$. (7) Based on the measurement outcomes from the previous step, Pauli-$X$ corrections are applied to every other qubit, starting from the fourth in the top row. The particular corrections needed are abbreviated by a multivariate function~$f$, the details of which are available in \eqref{eq:ghz-prep-correction}. (8) The measured qubits are reset to the all zeros state. (9) Final step of the preparation of the GHZ state of the control qubits. CNOT gates are again applied to every other control qubit. The final state of all control qubits is equal to a GHZ state. (10) Controlled-SWAPs are applied in parallel between control qubits and data qubits, in a first round of the implementation of the cyclic shift. (11) Controlled-SWAPs are applied in parallel between other control qubits and data qubits, in a second round of the implementation of the cyclic shift. (12) Hadamard gates are applied to all control qubits. (13) In a final step, all control qubits are measured in the computational basis and the measurement outcomes are processed according to \eqref{eq:real-part-estimator} to form an estimate of the real part of $\operatorname{Tr}[\rho_1 \cdots \rho_5]$. To estimate the imaginary part, replace every $H$ in step~(12) with $HS^{\dag}$. }%
\label{fig:2d-arch}%
\end{figure*}

We now outline how to implement our algorithm using a two-dimensional architecture similar to
Google's \cite{Arute2019}. We do so by means of a series of figures, which outline the time steps of the circuit implementation. See Figure~\ref{fig:2d-arch}. These figures can be understood as a two-dimensional implementation of the circuit depicted in Figure~\ref{fig:cyclic-shift-test}, with the exception that we also include qubit resets during the GHZ state preparation, as depicted in Figure~\ref{fig:ghz-prep-qubit-resets}. Explanations of the steps are given in the caption of Figure~\ref{fig:2d-arch}. The main point to highlight here is that our circuit leads to a highly parallelized implementation on a two-dimensional architecture  that should make it more amenable to realization on near-term quantum computers.

\section{Guarantees of our estimator}\label{sec:guarantees}

In this section, we show that the estimator $\hat{T}$ described in Section~\ref{subsec:explicit} is accurate and precise with high probability.

\begin{theorem}
\label{thm:estimate-trace-poly-func} Let $\{\rho_1, \ldots ,\rho_m\}$ be
single-qubit states. There exists a random variable $\hat{T}$ that can be computed with $O(\frac{1}{\varepsilon^2}\log(\frac{1}{\delta}))$ repetitions of a constant quantum depth circuit consisting of $O(m)$ three-qubit gates, and satisfies
\beq
\Pr(|\hat{T} - \operatorname{Tr}[\rho_1\cdots \rho_m]| \leq \varepsilon) \geq 1- \delta
\eeqp
\end{theorem}

\begin{proof}
By the Hoeffding inequality (see Lemma~\ref{lem:hoeffding}), it suffices to prove that the estimator $\hat{T}$ output by the method of Section~\ref{subsec:explicit} satisfies
\beq\label{eq:That} 
\mathbb{E}[\hat{T}] = \operatorname{Tr}[W^{\pi} (\rho_1\otimes\cdots \otimes \rho_m)] = \operatorname{Tr}[\rho_1\cdots \rho_m]
\eeqp

It suffices to prove the first equality. To begin with, let us suppose for simplicity that all the states are pure, i.e. $\rho_i = \ketbra{\psi_i}{\psi_i}$, and define the state of the target qubits as
\begin{equation}
|\psi^{(m)}\rangle \coloneqq  \ket{\psi_1} \otimes \cdots \otimes \ket{\psi_m}.
\end{equation}
Step 1 of our procedure prepares a GHZ\ state $|\Phi_{\text{GHZ}}^{\lfloor m/2 \rfloor}\rangle$ and adjoins it to $|\psi^{(m)}\rangle$, such that the overall state at the end of Step 1 is%
\begin{equation}
|\Phi_{\text{GHZ}}^{\lfloor m/2 \rfloor}\rangle|\psi^{(m)}\rangle.
\end{equation}
After Step 2 (the multiply-controlled cyclic shift), the
overall state becomes%
\begin{equation}
\frac{1}{\sqrt{2}}\left(  |0\rangle^{\otimes \lfloor m/2 \rfloor}|\psi^{(m)}\rangle
+|1\rangle^{\otimes \lfloor m/2 \rfloor}W^{\pi}|\psi^{(m)}\rangle\right)  .
\end{equation}
In Step 3, one measures the $\lfloor m/2 \rfloor$ control qubits in the $X$ basis, obtaining a $\lfloor m/2 \rfloor$ bits, such that the $i$-th bit is denoted by the random variable $X_i \in \{0,1\}$. We then compute the expectation of the random variable $\hat{R}$, defined as
\begin{equation}
\hat{R}\equiv \hat{R}(X_{1},\ldots,X_{\lfloor m/2 \rfloor})\coloneqq (-1)^{\sum_{i=1}^{\lfloor m/2 \rfloor}X_{i}},
\label{eq:real-part-estimator}
\end{equation}
which will be output as the real part of our estimator.
Introducing the following notation for $X$ basis eigenvectors
\begin{equation}
|\tilde{x}\rangle\coloneqq  \frac{1}{\sqrt{2}} \ket{0} + (-1)^x\ket{1} \quad \text{ for }x\in  \{0,1\},
\end{equation}
the probability of the $X$ basis measurement outputting the bitstring $x_1\cdots x_{\lfloor m/2 \rfloor}%
 \in \{0,1\}^{\lfloor m/2 \rfloor}$ is given
by 
\begin{align}
&  \Pr(x_1\cdots x_{\lfloor m/2 \rfloor})\nonumber\\
&  =\left\Vert
\begin{array}
[c]{c}%
\left(  \langle\widetilde{x_{1}},\widetilde{x_{2}},\cdots,\widetilde{x_{\lfloor m/2 \rfloor}%
}|\otimes I\right)  \times\\
\frac{1}{\sqrt{2}}\left(  |0\rangle^{\otimes \lfloor m/2 \rfloor}|\psi^{(m)}\rangle
+|1\rangle^{\otimes \lfloor m/2 \rfloor}W^{\pi}|\psi^{(m)}\rangle\right)
\end{array}
\right\Vert _{2}^{2} \label{eq:37}\\
&  =\frac{1}{2^{\lfloor m/2 \rfloor+1}}\left\Vert |\psi^{(m)}\rangle+(-1)^{\sum_{i=1}^{\lfloor m/2 \rfloor}x_{i}%
}W^{\pi}|\psi^{(m)}\rangle\right\Vert _{2}^{2}\\
&  =\frac{1+(-1)^{\sum_{i=1}^{\lfloor m/2 \rfloor}x_{i}}\operatorname{Re}[\operatorname{Tr}%
[W^{\pi}|\psi^{(m)}\rangle\!\langle\psi^{(m)}|]]}{2^{\lfloor m/2 \rfloor}}.
\end{align}
Thus, 
\begin{align}
&\mathbb{E}[\hat{R}] \nonumber \\
& =  \sum_{x_{1},\ldots,x_{\lfloor m/2 \rfloor}}\Pr(x_{1},\cdots,x_{\lfloor m/2 \rfloor})r(x_{1},\cdots,x_{\lfloor m/2 \rfloor}%
)\\
&  =\sum_{\substack{x_{1},\ldots,\\x_{\lfloor m/2 \rfloor}}} \!\!\! \left(  \frac{1+(-1)^{\sum_{i=1}^{\lfloor m/2 \rfloor}x_{i}%
}\operatorname{Re}[\operatorname{Tr}[W^{\pi}|\psi^{(m)}\rangle\!\langle
\psi^{(m)}|]]}{2^{\lfloor m/2 \rfloor}}\right)  \nonumber\\
& \hspace{150pt} \times (-1)^{\sum_{i=1}^{\lfloor m/2 \rfloor}x_{i}}\label{eq:expect-calc-real-part-1}\\
&  =\frac{1}{2^{\lfloor m/2 \rfloor}}\sum_{x_{1},\ldots,x_{\lfloor m/2 \rfloor}}\left(  (-1)^{\sum_{i=1}^{\lfloor m/2 \rfloor}x_{i}%
}\right.+ \nonumber\\
&\hspace{120pt} \left.\operatorname{Re}[\operatorname{Tr}[W^{\pi}|\psi^{(m)}\rangle\!\langle
\psi^{(m)}|]]\right)  \\
&  =\operatorname{Re}[\operatorname{Tr}[W^{\pi}|\psi^{(m)}\rangle\!\langle
\psi^{(m)}|]] , \label{eq:expect-calc-real-part-last}%
\end{align}
where, in the penultimate equality, we have used the fact that
\begin{equation}
\sum_{x_1,\ldots, x_\ell \in \{0,1\}^\ell} (-1)^{\sum_{i=1}^\ell x_i} = 0  \text{ for all } \ell \in \mathbb{N}. 
\end{equation}
The claim for mixed states $\rho^{(m)} \coloneqq \rho_1 \otimes \cdots \otimes \rho_m$, i.e., 
\beq\label{eq:44}
\mathbb{E}[\hat{R}] = \operatorname{Re}[\operatorname{Tr}[W^{\pi}\rho^{(m)}]],
\eeq
follows by convexity (i.e., that every mixed state
can be written as a convex combination of pure states). That is, we use the
fact that%
\begin{equation}
\Pr(x_{1},\ldots,x_{m})=\frac{1+(-1)^{\sum_{i=1}^{m}x_{i}}\operatorname{Re}%
[\operatorname{Tr}[W^{\pi}\rho^{(m)}]]}{2^{m}}%
\end{equation}
for this case and then repeat the calculation above for every pure state in the convex decomposition of $\rho^{(m)}$.

Using a similar chain of logic, we conclude that
\beq
\mathbb{E}[\hat{J}] = \operatorname{Im}[\operatorname{Tr}[W^{\pi}\rho^{(m)}]]
\eeqc
and the first equality of \eqref{eq:That} follows.
\end{proof}

\bigskip
We  also compute the variance of $\hat{T}$. Consider that $\operatorname{Var}[\hat{R}] \coloneqq  \mathbb{E}[(\hat{R}-\mathbb{E}[\hat{R}])^{2}]=\mathbb{E}[\hat{R}^{2}]-\mathbb{E}[\hat{R}]^{2}$. Since 
\begin{align}
\mathbb{E}[\hat{R}^{2}]  & =\sum_{x_{1},\cdots,x_{m}}\Pr(x_{1},\cdots,x_{m})\left(
r(x_{1},\cdots,x_{m})\right)  ^{2}\label{eq:variance-calc-1}\\
& =\sum_{x_{1},\cdots,x_{m}}\frac{1+(-1)^{\sum_{i=1}^{m}x_{i}}%
\operatorname{Re}[\operatorname{Tr}[W^{\pi}\rho^{(m)}]]}{2^{m}}\\
& =1,\label{eq:variance-calc-3}%
\end{align}
we conclude that $\operatorname{Var}[\hat{R}] = 1-(\operatorname{Re}
[\operatorname{Tr}[W^{\pi}\rho^{(m)}]])^{2}$. Similarly, $\operatorname{Var}[\hat{J}] = 1-(\operatorname{Im}
[\operatorname{Tr}[W^{\pi}\rho^{(m)}]])^{2}.$ Since these two random variables are independent, 
\beq
\operatorname{Var}[\hat{T}] = 2- \left|\operatorname{Tr}[W^{\pi}\rho^{(m)}]\right|^2
\eeqp

We now discuss the generalization of
Theorem~\ref{thm:estimate-trace-poly-func}\ to states of more than one
qubit. This generalization can be accomplished by increasing the circuit width
or the circuit depth, as discussed in Section~\ref{subsec:explicit}.

\begin{proposition}
\label{prop:estimate-trace-poly-func-2} 
Let $\{\rho_1, \ldots ,\rho_m\}$ be  a set of 
$p$-qubit states, and fix $\varepsilon >0$ and $\delta \in (0,1)$. There exists a random variable $\hat{T}_p$ that can be computed using $O(\frac{1}{\varepsilon^2}\log(\frac{1}{\delta}))$ repetitions of a constant quantum depth circuit consisting of $O(mp)$ three-qubit gates, and satisfies
\beq\label{eq:Tp}
\Pr(|\hat{T}_p - \operatorname{Tr}[\rho_1\cdots \rho_m]| \leq \varepsilon) \geq 1- \delta
\eeqp
\end{proposition}

\begin{proof}
The gate count for the circuits for both constructions detailed in Section~\ref{subsec:explicit} is $O(mp)$. For the second construction (increasing the depth), Eq.~\eqref{eq:Tp} follows from the exact same calculation as in the proof of Theorem \ref{thm:estimate-trace-poly-func}. For the first construction (increasing the width), let us define $\rho^{(m,p)} \coloneqq  \rho_1 \otimes \cdots \otimes \rho_m.$ It suffices to prove that for the estimator $\hat{R}_p \coloneqq \left(  -1\right)  ^{X_{1}+\cdots+X_{mp}}$ derived from the $mp$ measurement outcomes $X_{1}, \cdots, X_{mp}$ fulfils $\mathbb{E}[\hat{R}_p ]=\operatorname{Re}[\operatorname{Tr}[W^{\pi}%
\rho^{(m,p)}]]$. This follows from a  calculation similar to that in \eqref{eq:37}--\eqref{eq:44}.
\end{proof}

\bigskip
Similarly, 
\beq
\operatorname{Var}[\hat{T}_p] = 2- \left|\operatorname{Tr}[W^{\pi}\rho^{(m,p)}]\right|^2
\eeqp

\section{Applications of our method} \label{sec:app}

An application of our method is to estimate functions of density matrices that can be approximated by `well-behaved' polynomials. This was already suggested in the original work of \cite{EAOHHK02}, but no complexity analysis was put forth. Here we formalize the analysis of the application in the following theorem:

\begin{theorem}
\label{thm:gen-app}
Let $\rho$ be a quantum state with rank at most~$d$. 
Suppose there exists a constant $ \varepsilon > 0$ and a function $C$ such that $g:\mathbb{R} \rightarrow \mathbb{R}$ is approximated by a degree-$m$ polynomial $f(x) = \sum_{k=0}^m c_k x^k$ on the interval $[0,1]$, in the sense that
\beq\label{eq:approxerror}
\sup_{x \in [0,1]} \left| g(x) - f(x) \right| < \frac{\varepsilon}{2d}
\eeqc
and  
\beq\label{eq:bounded_coeffs}
 \sum_{k=0}^m \left|c_k\right| < C
\eeqp
Then estimating 
$
\Tr[g(\rho)]
$
within $\varepsilon$ additive error and with success probability not smaller than $1-\delta$
requires $O(m^2 \frac{C^2}{\varepsilon^2} \log(\frac{1}{\delta}))$ copies of $\rho$ and $O(m\frac{C^2}{\varepsilon^2} \log(\frac{1}{\delta}))$ runs of a circuit with $O(m)$ controlled SWAP gates. 
\end{theorem}
(Here, we should think of $C$ as a slowly-growing function of $m$.)

An example of a function of a quantum state that can be estimated in this way is $g(x) = \left(1+x\right)^{\alpha}$ for $\alpha>0$, which has the following expansion as a binomial series
\begin{equation}
(1+x)^\alpha = \sum_{k=0}^\infty \binom{\alpha}{k} x^k , 
\end{equation}
where $\binom{\alpha}{k}$ is the generalized binomial coefficient. It is well known that the binomial series converges absolutely for $x=1$ and $\alpha > 0$, which implies that, for every $\alpha > 0$, there exists a positive constant $C(\alpha)$ such that
\begin{equation}
\sum_{k=0}^m \left|\binom{\alpha}{k}\right| \leq \sum_{k=0}^\infty \left|\binom{\alpha}{k}\right| = C(\alpha).
\end{equation}
Thus, this series satisfies the criterion in \eqref{eq:bounded_coeffs}.

Another function of a quantum state that can be estimated in this way is $g(x)=\ln(x+1)$. Indeed, this function has the following well known expansion:
\begin{equation}
\ln(x+1) = \sum_{k=1}^\infty \frac{(-1)^k}{k} x^k,
\end{equation}
so that the absolute partial sum of the coefficients satisfies $\sum_{k=1}^m \left|\frac{(-1)^k}{k}\right| \approx \ln m + \gamma$, where $\gamma \approx 0.577$ is the Euler--Mascheroni constant. The point here is that, even though the absolute partial sums are not bounded by a constant, they still grow sufficiently slowly such that the algorithm runs efficiently.

Yet another example of a function, but in this case with applications in thermodynamics, is $g(x) = e^{\beta x}$, where $\beta \in \mathbb{R}$. Here we have that
\begin{equation}
    e^{\beta x} = \sum_{k=0}^{\infty} \frac{(\beta x)^k}{k!},
\end{equation}
so that $c_k = \beta^k / k!$ and thus
\begin{equation}
    \sum_{k=0}^m |c_k| \leq \sum_{k=0}^\infty \frac{|\beta|^k}{k!} = e^{|\beta|}.
\end{equation}
Thus, in this case the absolute partial sums can be bounded by a constant.
This function is particularly interesting for thermodynamics applications (and possibly beyond) because our method allows for estimating the partition function $\operatorname{Tr}[e^{\beta \rho}]$ whenever the Hamiltonian is encoded in a density matrix $\rho$, as assumed, e.g., in the density matrix exponentiation algorithm \cite{Lloyd2014QuantumAnalysis,Kimmel2017HamiltonianComplexity}.
\bigskip

\begin{proof}[Proof of Theorem~\ref{thm:gen-app}]
We will present an estimator for the desired quantity that satisfies the claimed complexity guarantees. Let us describe the estimator for $\operatorname{Re}[\Tr[g(\rho)]]$ (the estimator for the imaginary part follows immediately). To estimate $\operatorname{Re}[\Tr[g(\rho)]]$, we run the following procedure:
\begin{enumerate}
\item For each $k\in [m]$, run the circuit described in Steps~1-3 of Section~\ref{subsec:explicit} once to output a random variable $R_k \in \{-1,1\}$ such that 
\beq
\mathbb{E}[R_k] = \operatorname{Re}[\Tr[\rho^k]]
\eeqp

\item Linearly combine the above random variables to form the new random variable
\beq\label{eq:Rl}
R^{(1)} = \sum_{k=0}^m c_k R_k
\eeqp

\item Let 
\begin{equation}
N = \left \lceil \frac{8C^{2}}{\varepsilon^{2}}\ln\!\left(  \frac{2}{\delta}\right)\right \rceil .  
\end{equation}
Repeat the first two steps $N-1$ more times, on the $i$-th iteration outputting the random variable $R^{(i)}$. 

\item Output $\hat{g} = \frac{1}{N} \sum_{i=1}^N R^{(i)}$ as the estimator for $\operatorname{Re}[\Tr[g(\rho)]]$. 
\end{enumerate}

Let us now prove the correctness of this procedure. Suppose the spectral decomposition of $\rho$ is as follows:
\beq
\rho = \sum_{i=1}^{r_{\rho}} \lambda_i \ketbra{\psi_i}{\psi_i}
\eeqc
where $r_{\rho}$ is the rank of $\rho$. In the limit of $N \rightarrow \infty$, the only error in the estimator $\hat{g}$ would come from the error in the polynomial approximation. That is, 
\begin{align}
| \Tr[g(\rho)] - \sum_{k=0}^m c_k \Tr[\rho^k]| & = \left|\sum_{i=1}^{r_{\rho}} \left(g(\lambda_i) - \sum_{k=0}^m c_k \lambda_i^k \right)\right| \notag \\& \leq \varepsilon/2 \, ,
\label{eq:apperr}
\end{align}
where the inequality follows from \eqref{eq:approxerror} and the fact that $d > r_{\rho}$.  Now we account for the other source of error, which is the statistical error caused by taking a finite number of samples. Consider that
\begin{align}
|R^{(i)}| \leq \sum_{k=0}^m |c_k| |R_k| \leq \sum_{k=0}^m |c_k| \leq C \, ,
\label{eq:bnd-RV-for-hoeffding}
\end{align}
for all $i\in \{1,\ldots, N\}$,
where the first inequality follows from the triangle inequality, the second from the fact that $R_k \in \{-1,1\}$, and the third from the assumption in \eqref{eq:bounded_coeffs}.
By applying \eqref{eq:bnd-RV-for-hoeffding}, the Hoeffding inequality in Lemma~\ref{lem:hoeffding} immediately applies to our setting when we take $Y_i \gets R^{(i)}$ and $[a,b] \gets [-C,C]$. We also see that $\mu \gets \mathbb{E}[ R^{(1)}] = \sum_{k=0}^m c_k \Tr[\rho^k]$. Then if we set $N$ to 
\begin{equation}
N = \left \lceil \frac{8C^{2}}{\varepsilon^{2}}\ln\!\left(  \frac{2}{\delta}\right)  \right \rceil \, ,
\end{equation}
we get that 
\begin{equation}\label{eq:staterr}
\Pr\!\left[  \left\vert \hat{g} -\sum_{k=0}^m c_k \Tr[\rho^k] \right\vert \leq \varepsilon/2
\right]  \geq1-\delta \, .
\end{equation}
Combining the two sources of error in \eqref{eq:apperr} and \eqref{eq:staterr}, we get that with probability $1-\delta$, 
\begin{multline}
 \left\vert \hat{g} -\operatorname{Re}[\Tr[g(\rho)]] \right\vert 
 \leq \left\vert \hat{g} -\sum_{k=0}^m c_k \Tr[\rho^k] \right\vert + \\\left\vert\operatorname{Re}[\Tr[g(\rho)]]  - \sum_{k=0}^m c_k \Tr[\rho^k] \right\vert 
 \leq \varepsilon \, .
\end{multline}
Repeating the analysis for the estimation of $\operatorname{Im}[\Tr[g(\rho)]]$ yields the stated complexity.
\end{proof}

\section{Discussion}

An alternative way to estimate $\Tr[\rho^k]$ for each $k \in [m]$ is by using single-copy randomized measurements to estimate nonlinear functions of a density matrix \cite{EB12}. More recent approaches use the method of classical shadows to obtain `classical snapshots' of $\rho$ that can be linearly combined to obtain a classical random variable whose expectation is $\Tr[\rho^k]$ (see \cite[Supplementary Material Section~6]{Huang20} and \cite{RBMV2021}). The latter reference provides rigorous bounds on the sample complexity required for the estimation of quantities like $\Tr(O^{q}\rho^{\otimes q})$ for arbitrary $q$, where $O^{q}$ is an observable acting on the Hilbert space of $q$ systems; setting $O^{q}$ to be the real and imaginary parts of $W^{\pi}$ on the $q$ systems and taking their linear combination, we recover our setting. 

The benefit of these alternative methods is that they allow for sequential measurements and do not require the assumption that the copies of $\rho$ available to the algorithm are identical and independent. However, this robustness comes at a significant cost in terms of sample and computational complexity, as they generically require a number of copies scaling in the dimension $d$ of the states involved -- i.e. exponential in $n$ -- as well as poly$(1/\delta)$ where $\delta$ is the failure probability of the estimation (see \cite[Appendix E]{RBMV2021}). Computational complexity is always lower-bounded by sample complexity. On the other hand, our sample complexity does not depend on $d$ at all, and the dependence on $1/\delta$ is only logarithmic. However, our {\em computational} complexity does depend on $d$ as we use $d$-dimensional SWAP gates (but these have complexity $O(\log d)$).

One might also wonder whether we could use this approach to estimate R\'enyi or von Neumann entropies of quantum states. The main difficulty in doing so is that well known polynomial approximations of the functions $x^\alpha$ and $-x \ln x$, given respectively by
\begin{align}
x^\alpha & = \sum_{k=0}^\infty \binom{\alpha}{k} (x-1)^k \\
& = \sum_{k=0}^\infty \binom{\alpha}{k} \sum_{\ell = 0 }^k \binom{k}{\ell}(-1)^{k-\ell} x^\ell \, ,\\
-x \ln x & = x\sum_{k=1}^\infty \frac{(1-x)^k}{k}\\
& = x\sum_{k=1}^\infty \frac{1}{k} \sum_{\ell=0}^k \binom{k}{\ell} (-1)^\ell x^\ell \\
& = \sum_{k=1}^\infty \frac{1}{k} \sum_{\ell=0}^k \binom{k}{\ell} (-1)^\ell x^{\ell+1} \, ,
\end{align}
do not satisfy the condition in \eqref{eq:bounded_coeffs}, in the sense that the absolute partial sums grow too quickly and therefore do not lead to an efficient algorithm using this approach. See also \cite{stack21} in this context. It thus remains open whether this approach can be used effectively for estimating these important uncertainty measures. See \cite{JVHW15,wu2016minimax,JVHW17,AOST17} for work on this topic in classical information theory and \cite{AISW20,GL20,LS20,SH21,WZW22,gur2021sublinear,li2022unified} for a flurry of recent efforts on estimating R\'enyi and von Neumann entropies using quantum computers, which propose alternative approaches. The authors of \cite{wang_fidelity_2023, gilyén2022improved} have also proposed and analyzed quantum algorithms for fidelity estimation.

We note here that the method outlined above can be generalized to functions of multiple density matrices. For example, let $\rho$ and $\sigma$ be quantum states, and suppose that $g_1$ and $g_2$ are well behaved polynomials in the sense described in Theorem~\ref{thm:gen-app}. Then we can employ a similar approach to estimate the functions $\Tr[g_1(\rho)g_2(\sigma)]$ and $\Tr[g_1(\sigma^{1/2}\rho\sigma^{1/2})]$. Polynomial approximations of these functions take the following form:
\begin{align}
  \Tr\!\left[\left(\sum_{k} c_k\rho^k\right)\left( \sum_{\ell}d_\ell \sigma^\ell\right)\right] & = \sum_{k,\ell} c_k d_\ell \Tr[\rho^k\sigma^\ell], \notag \\
\sum_{k} a_k \Tr[(\sigma^{1/2}\rho\sigma^{1/2})^k] &= \sum_{k} a_k \Tr[(\rho\sigma)^k], 
\end{align}
respectively, and can be estimated using our circuits combined with classical postprocessing.
Thus, by the discussion after Theorem~\ref{thm:gen-app}, we can take $g_1(x) = (1+x)^\alpha$ and $g_2(x)=(1+x)^\beta$ for $\alpha,\beta > 0$. A case of interest is when we set $\alpha \in (0,1)$ and $\beta = 1-\alpha$. The resulting function $\Tr[g_1(\rho)g_2(\sigma)]$ then satisfies faithfulness and the data-processing inequality under unital quantum channels~\cite{PWPR06} and thus can serve as an alternative to the widely used Hilbert--Schmidt distance measure (which also satisfies the data-processing inequality under unital channels~\cite{PWPR06}). We prove these claims in Appendix~\ref{app:dp-proof}.

Other functions of two density matrices of interest include the Schatten-$p$ distances. For states $\rho$ and $\sigma$, these are defined for $p>1$ as
\begin{equation}
\left \Vert \rho - \sigma \right \Vert_p \coloneqq \left(\operatorname{Tr}\left[|\rho -\sigma|^p\right]\right)^{\frac{1}{p}}.
\end{equation}
The simplest of these is the Hilbert--Schmidt distance (i.e., when $p=2$), which reduces to
\begin{equation}
\left \Vert \rho - \sigma \right \Vert_2^2 = \operatorname{Tr}[\rho^2] + \operatorname{Tr}[\sigma^2] - 2 \operatorname{Tr}[\rho\sigma].
\end{equation}
This distance measure has been extensively employed in various works on quantum computing, with applications to compiling and machine learning, mainly because it is convenient to estimate it on quantum computers using the well known swap test. We can also consider the cases of $p=4$ and $p=6$, leading to the following formulas proved in Appendix~\ref{app:schatten-dist-proofs}:
\begin{align}
\left \Vert \rho - \sigma \right \Vert_4^4 & = \operatorname{Tr}[\rho^{4}]-4\operatorname{Tr}[\rho^{3}\sigma
]+4\operatorname{Tr}[\rho^{2}\sigma^{2}]\notag \\
& \quad +2\operatorname{Tr}[\rho\sigma
\rho\sigma]-4\operatorname{Tr}[\rho\sigma^{3}]+\operatorname{Tr}[\sigma^{4}],\\
\left \Vert \rho - \sigma \right \Vert_6^6 & = \operatorname{Tr}[\rho^{6}]-6\operatorname{Tr}[\rho^{5}\sigma
]+6\operatorname{Tr}[\rho^{4}\sigma^{2}] \notag \\
& \quad +6\operatorname{Tr}[\rho^{3}\sigma
\rho\sigma]+3\operatorname{Tr}[\rho^{2}\sigma\rho^{2}\sigma
]\notag \\
& \quad -6\operatorname{Tr}[\rho^{2}\sigma^{2}\rho\sigma] -6\operatorname{Tr}[\rho^{3}\sigma^{3}]\notag \\
& \quad-3\operatorname{Tr}[\rho^{2}%
\sigma\rho\sigma^{2}] -2\operatorname{Tr}[\rho\sigma\rho\sigma\rho
\sigma]\notag \\
& \quad+6\operatorname{Tr}[\rho\sigma\rho\sigma^{3}]-3\operatorname{Tr}%
[\rho^{2}\sigma\rho\sigma^{2}]\notag \\
&\quad  +3\operatorname{Tr}[\rho\sigma^{2}\rho\sigma^{2}]+6\operatorname{Tr}%
[\rho^{2}\sigma^{4}]\notag \\
& \quad -6\operatorname{Tr}[\rho\sigma^{5}]+\operatorname{Tr}%
[\sigma^{6}].
\label{eq:schatten-6}
\end{align}
These distances can also be estimated using our algorithms proposed here, because each of the terms involved is a multivariate trace. It remains open for future work to use these distances in applications like compiling and machine learning.



\section{Conclusion}

We have provided a quantum circuit for multivariate trace estimation that requires only constant quantum depth, and hence it is more amenable to be implemented on near-term quantum computers than previous methods that require linear depth. Our architecture is also flexible and can be smoothly tailored to the availability of circuit width (at the cost of more-entangling gates). 

Going forward from here, one can further consider the application of our method to estimating nonlinear functions of quantum states. \textquotedblleft The most important application of
computers has been designing better computers\textquotedblright\ \cite{V19}, and these methods can be used for this purpose. Our method to estimate functions of quantum states based on their polynomial approximations opens the door to the idea that {\em near-term} quantum computers can be used to design better quantum computers. An important open question in this regard is whether any functions whose polynomial approximations fulfill \eqref{eq:bounded_coeffs} have an interpretation as quality metrics for quantum computer design. Conversely, it would also be interesting to explore further whether there are any quantum state distinguishability measures  -- critical in applications like quantum compiling
\cite{Khatri2019quantumassisted,SKCC20}\ and state learning
\cite{LLB18,CSZW20} -- that fulfill this condition. 

Our findings open up many avenues for future research. While we already achieve the optimal possible circuit depth, other aspects like circuit width, gate number, and noise robustness could be further optimized to fit near-term constraints. See Ref.~\cite{Liang_trace_estimation_2023} for some recent developments in these directions.

A possible extension for future work is to combine our algorithm with the crosstalk mitigation techniques of Ref.~\cite{cross_talk_miti} to make them directly applicable to noisy quantum processors. As mentioned before, a related exploration was performed recently in \cite{Liang_trace_estimation_2023}, wherein the authors combined our approach with error mitigation and explored circuit depth-width trade-offs. Another possible avenue is to employ amplitude estimation techniques \cite{Montanaro2015} to improve the dependence of our algorithm on $\varepsilon$ from $\frac{1}{\varepsilon^2}$ to $\frac{1}{\varepsilon}$; however, doing so using known techniques will incur a significant increase in circuit depth~\cite{GiurgicaTiron2022lowdepthalgorithms}, and thus we would find it surprising if a constant quantum depth circuit could have this dependence on the accuracy parameter~$\varepsilon$. Heuristic methods have been explored in \cite{Plekhanov2022variationalquantum} and can also be considered for this purpose. 

%

\medskip
\textbf{Acknowledgments}.~We acknowledge helpful discussions with Jayadev Acharya, Antonio Anna Mele, Patrick Coles,
Andr\'as Gily\'en, Zo\"{e} Holmes, Dhrumil Patel, Eliott Rosenberg, Aliza
Siddiqui, and Kevin Valson Jacob. We also acknowledge helpful comments from the anonymous referees, which improved our paper. YQ\ acknowledges support from a Stanford QFARM
fellowship, an NUS Overseas Graduate Scholarship, and an Alexander von Humboldt Fellowship. EK and MMW\ acknowledge support from the National Science Foundation
(NSF)\ under Grant No.~1714215. Part of this work was conducted while EK was employed at the Department of Physics and Astronomy of Louisiana
State University.

\bibliographystyle{quantum}
\bibliography{Ref}

\appendix

\section{Proofs of faithfulness and data processing}

\label{app:dp-proof}

Let us define the following measures for states $\rho$ and $\sigma$ and
$\alpha\in\left(  0,1\right)  \cup\left(  1,\infty\right)  $:%
\begin{align}
K_{\alpha}(\rho\Vert\sigma)  & \coloneqq \operatorname{Tr}[\left(  I+\rho\right)
^{\alpha}\left(  I+\sigma\right)  ^{1-\alpha}],\\
Q_{\alpha}(\rho\Vert\sigma)  & \coloneqq \operatorname{Tr}[\rho^{\alpha}%
\sigma^{1-\alpha}].
\end{align}
These measures are related as follows:%
\begin{equation}
K_{\alpha}(\rho\Vert\sigma)=\left(  d+1\right)  Q_{\alpha}\!\left(
\frac{I+\rho}{d+1}\middle\Vert\frac{I+\sigma}{d+1}\right)  ,\label{eq:K-to-Q}%
\end{equation}
where we observe that $\frac{I+\rho}{d+1}$ and $\frac{I+\sigma}{d+1}$ are states.
It is known that%
\begin{equation}
0\leq Q_{\alpha}(\rho\Vert\sigma)\leq1
\end{equation}
for all states $\rho$ and $\sigma$ (the lower bound follows because $\rho$ and
$\sigma$ are positive semi-definite and the upper bound follows by applying
the H\"{o}lder inequality). Furthermore, the measure $Q_{\alpha}(\rho
\Vert\sigma)$ is faithful on states, i.e., equal to~$1$ if and only if
$\rho=\sigma$, and it satisfies the data-processing inequality \cite{P85,P86}:%
\begin{align}
Q_{\alpha}(\rho\Vert\sigma)  & \leq Q_{\alpha}(\mathcal{N}(\rho)\Vert
\mathcal{N}(\sigma)),\quad\text{for }\alpha\in\left(  0,1\right)  ,\\
Q_{\alpha}(\rho\Vert\sigma)  & \geq Q_{\alpha}(\mathcal{N}(\rho)\Vert
\mathcal{N}(\sigma)),\quad\text{for }\alpha\in(1,2],\label{eq:DP-Q-a-1-2}%
\end{align}
for every channel $\mathcal{N}$.

By the equality in \eqref{eq:K-to-Q}, we can conclude properties of
$K_{\alpha}(\rho\Vert\sigma)$ from properties of $Q_{\alpha}(\rho\Vert\sigma
)$. Indeed,%
\begin{equation}
0\leq K_{\alpha}(\rho\Vert\sigma)\leq d+1.
\end{equation}
Also, the measure $K_{\alpha}(\rho\Vert\sigma)$\ is faithful, i.e., equal to
$d+1$ if and only if $\rho=\sigma$. To see this, consider that%
\begin{equation}
Q_{\alpha}\!\left(  \frac{I+\rho}{d+1}\middle\Vert\frac{I+\sigma}{d+1}\right)
=1
\end{equation}
if and only if
\begin{equation}
\frac{I+\rho}{d+1}=\frac{I+\sigma}{d+1}.    
\end{equation}
This last equality
is equivalent to $\rho=\sigma$. Thus, the faithfulness claim follows. Finally,
the measure $K_{\alpha}(\rho\Vert\sigma)$ obeys the data-processing inequality
for unital quantum channels. For $\alpha\in\left(  0,1\right)  $ and a unital
channel~$\mathcal{N}$ (i.e., satisfying $\mathcal{N}(I)=I$), we have that
\begin{equation}
K_{\alpha}(\rho\Vert\sigma)\leq K_{\alpha}(\mathcal{N}(\rho)\Vert
\mathcal{N}(\sigma)),
\end{equation}
and for $\alpha\in(1,2]$, we have that%
\begin{equation}
K_{\alpha}(\rho\Vert\sigma)\geq K_{\alpha}(\mathcal{N}(\rho)\Vert
\mathcal{N}(\sigma)).
\end{equation}
These inequalities follow from the data-processing inequality for $Q_{\alpha}%
$. Indeed, consider for $\alpha\in\left(  0,1\right)  $ that
\begin{align}
& K_{\alpha}(\rho\Vert\sigma) \notag \\
&  =\left(  d+1\right)  Q_{\alpha}\!\left(
\frac{I+\rho}{d+1}\middle\Vert\frac{I+\sigma}{d+1}\right)  \\
&  \leq\left(  d+1\right)  Q_{\alpha}\!\left(  \mathcal{N}\left(  \frac
{I+\rho}{d+1}\right)  \middle\Vert\mathcal{N}\left(  \frac{I+\sigma}%
{d+1}\right)  \right)  \\
&  =\left(  d+1\right)  Q_{\alpha}\!\left(  \frac{I+\mathcal{N}(\rho)}%
{d+1}\middle\Vert\frac{I+\mathcal{N}(\sigma)}{d+1}\right)  \\
&  =K_{\alpha}(\mathcal{N}(\rho)\Vert\mathcal{N}(\sigma)).
\end{align}
The second equality follows from linearity of the channel~$\mathcal{N}$ and
the fact that it is unital. The inequality for $\alpha\in(1,2]$ follows
similar reasoning as above but instead makes use of \eqref{eq:DP-Q-a-1-2}.

\section{Schatten \texorpdfstring{$p$}{p}-distances as linear combinations of multivariate traces}

\label{app:schatten-dist-proofs}

In this appendix, we establish formulas for the Schatten-4 and -6 distances of
quantum states, in terms of linear combinations of multivariate traces. To
start, let us define the Schatten-$2k$ distance measure of states $\rho$ and
$\sigma$ as follows:%
\begin{align}
\left\Vert \rho-\sigma\right\Vert _{2k} &  \coloneqq \operatorname{Tr}[\left\vert
\rho-\sigma\right\vert ^{2k}]^{\frac{1}{2k}}\\
&  =\operatorname{Tr}[\left(  \rho-\sigma\right)  ^{2k}]^{\frac{1}{2k}},
\end{align}
where $k\in\mathbb{Z}^{+}$. Let us consider the cases $k\in\left\{
1,2,3\right\}  $. First consider $k=1$:
\begin{align}
\left(  \rho-\sigma\right)  ^{2} & =\left(  \rho-\sigma\right)  \left(
\rho-\sigma\right)  \\
& =\rho^{2}-\rho\sigma-\sigma\rho+\sigma^{2},
\end{align}
which implies that%
\begin{equation}
\operatorname{Tr}[\left(  \rho-\sigma\right)  ^{2}]=\operatorname{Tr}[\rho
^{2}]-2\operatorname{Tr}[\rho\sigma]+\operatorname{Tr}[\sigma^{2}],
\end{equation}
giving that%
\begin{equation}
\left\Vert \rho-\sigma\right\Vert _{2}^{2}=\operatorname{Tr}[\rho
^{2}]-2\operatorname{Tr}[\rho\sigma]+\operatorname{Tr}[\sigma^{2}].
\end{equation}
This is the well known Hilbert--Schmidt distance measure.

\vspace{.5in}

\onecolumngrid
\hrulefill 

\vspace{.5in}

Now consider $k=2$:%
\begin{align}
\left(  \rho-\sigma\right)  ^{4} &  =\left(  \rho-\sigma\right)  ^{2}\left(
\rho-\sigma\right)  ^{2}\\
&  =\left(  \rho^{2}-\rho\sigma-\sigma\rho+\sigma^{2}\right)  \left(  \rho
^{2}-\rho\sigma-\sigma\rho+\sigma^{2}\right)  \\
&  =\left(  \rho^{2}-\rho\sigma-\sigma\rho+\sigma^{2}\right)  \rho^{2}-\left(
\rho^{2}-\rho\sigma-\sigma\rho+\sigma^{2}\right)  \rho\sigma\nonumber\\
&  \quad-\left(  \rho^{2}-\rho\sigma-\sigma\rho+\sigma^{2}\right)  \sigma
\rho+\left(  \rho^{2}-\rho\sigma-\sigma\rho+\sigma^{2}\right)  \sigma^{2}\\
&  =\rho^{2}\rho^{2}-\rho\sigma\rho^{2}-\sigma\rho\rho^{2}+\sigma^{2}\rho
^{2}-\rho^{2}\rho\sigma+\rho\sigma\rho\sigma+\sigma\rho\rho\sigma-\sigma
^{2}\rho\sigma\nonumber\\
&  \quad-\rho^{2}\sigma\rho+\rho\sigma\sigma\rho+\sigma\rho\sigma\rho
-\sigma^{2}\sigma\rho+\rho^{2}\sigma^{2}-\rho\sigma\sigma^{2}-\sigma\rho
\sigma^{2}+\sigma^{2}\sigma^{2}\\
&  =\rho^{4}-\rho\sigma\rho^{2}-\sigma\rho^{3}+\sigma^{2}\rho^{2}-\rho
^{3}\sigma+\rho\sigma\rho\sigma+\sigma\rho^{2}\sigma-\sigma^{2}\rho
\sigma\nonumber\\
&  \quad-\rho^{2}\sigma\rho+\rho\sigma^{2}\rho+\sigma\rho\sigma\rho-\sigma
^{3}\rho+\rho^{2}\sigma^{2}-\rho\sigma^{3}-\sigma\rho\sigma^{2}+\sigma^{4}.
\end{align}
Now taking the trace, we find that%
\begin{align}
\operatorname{Tr}[\left(  \rho-\sigma\right)  ^{4}] &  =\operatorname{Tr}%
[\rho^{4}]-\operatorname{Tr}[\rho\sigma\rho^{2}]-\operatorname{Tr}[\sigma
\rho^{3}]+\operatorname{Tr}[\sigma^{2}\rho^{2}]-\operatorname{Tr}[\rho
^{3}\sigma]+\operatorname{Tr}[\rho\sigma\rho\sigma]+\operatorname{Tr}%
[\sigma\rho^{2}\sigma]-\operatorname{Tr}[\sigma^{2}\rho\sigma]\nonumber\\
&  \quad-\operatorname{Tr}[\rho^{2}\sigma\rho]+\operatorname{Tr}[\rho
\sigma^{2}\rho]+\operatorname{Tr}[\sigma\rho\sigma\rho]-\operatorname{Tr}%
[\sigma^{3}\rho]+\operatorname{Tr}[\rho^{2}\sigma^{2}]-\operatorname{Tr}%
[\rho\sigma^{3}]-\operatorname{Tr}[\sigma\rho\sigma^{2}]+\operatorname{Tr}%
[\sigma^{4}]\\
&  =\operatorname{Tr}[\rho^{4}]-\operatorname{Tr}[\rho^{3}\sigma
]-\operatorname{Tr}[\rho^{3}\sigma]+\operatorname{Tr}[\rho^{2}\sigma
^{2}]-\operatorname{Tr}[\rho^{3}\sigma]+\operatorname{Tr}[\rho\sigma\rho
\sigma]+\operatorname{Tr}[\rho^{2}\sigma^{2}]-\operatorname{Tr}[\rho\sigma
^{3}]\nonumber\\
&  \quad-\operatorname{Tr}[\rho^{3}\sigma]+\operatorname{Tr}[\rho^{2}%
\sigma^{2}]+\operatorname{Tr}[\rho\sigma\rho\sigma]-\operatorname{Tr}%
[\rho\sigma^{3}]+\operatorname{Tr}[\rho^{2}\sigma^{2}]-\operatorname{Tr}%
[\rho\sigma^{3}]-\operatorname{Tr}[\rho\sigma^{3}]+\operatorname{Tr}%
[\sigma^{4}]\\
&  =\operatorname{Tr}[\rho^{4}]-4\operatorname{Tr}[\rho^{3}\sigma
]+4\operatorname{Tr}[\rho^{2}\sigma^{2}]+2\operatorname{Tr}[\rho\sigma
\rho\sigma]-4\operatorname{Tr}[\rho\sigma^{3}]+\operatorname{Tr}[\sigma^{4}].
\end{align}

Let us continue and consider the case of $k=3$:%
\begin{align}
\left(  \rho-\sigma\right)  ^{6} &  =\left(  \rho-\sigma\right)  ^{4}\left(
\rho-\sigma\right)  ^{2}\\
&  =(\rho^{4}-\rho\sigma\rho^{2}-\sigma\rho^{3}+\sigma^{2}\rho^{2}-\rho
^{3}\sigma+\rho\sigma\rho\sigma+\sigma\rho^{2}\sigma-\sigma^{2}\rho
\sigma\nonumber\\
&  \quad-\rho^{2}\sigma\rho+\rho\sigma^{2}\rho+\sigma\rho\sigma\rho-\sigma
^{3}\rho+\rho^{2}\sigma^{2}-\rho\sigma^{3}-\sigma\rho\sigma^{2}+\sigma
^{4})\left(  \rho-\sigma\right)  ^{2}\\
&  =(\rho^{4}-\rho\sigma\rho^{2}-\sigma\rho^{3}+\sigma^{2}\rho^{2}-\rho
^{3}\sigma+\rho\sigma\rho\sigma+\sigma\rho^{2}\sigma-\sigma^{2}\rho
\sigma\nonumber\\
&  \quad-\rho^{2}\sigma\rho+\rho\sigma^{2}\rho+\sigma\rho\sigma\rho-\sigma
^{3}\rho+\rho^{2}\sigma^{2}-\rho\sigma^{3}-\sigma\rho\sigma^{2}+\sigma
^{4})\left(  \rho^{2}-\rho\sigma-\sigma\rho+\sigma^{2}\right)  \\
&  =(\rho^{4}-\rho\sigma\rho^{2}-\sigma\rho^{3}+\sigma^{2}\rho^{2}-\rho
^{3}\sigma+\rho\sigma\rho\sigma+\sigma\rho^{2}\sigma-\sigma^{2}\rho
\sigma\nonumber\\
&  \quad-\rho^{2}\sigma\rho+\rho\sigma^{2}\rho+\sigma\rho\sigma\rho-\sigma
^{3}\rho+\rho^{2}\sigma^{2}-\rho\sigma^{3}-\sigma\rho\sigma^{2}+\sigma
^{4})\rho^{2}\nonumber\\
&  \quad-(\rho^{4}-\rho\sigma\rho^{2}-\sigma\rho^{3}+\sigma^{2}\rho^{2}%
-\rho^{3}\sigma+\rho\sigma\rho\sigma+\sigma\rho^{2}\sigma-\sigma^{2}\rho
\sigma\nonumber\\
&  \quad-\rho^{2}\sigma\rho+\rho\sigma^{2}\rho+\sigma\rho\sigma\rho-\sigma
^{3}\rho+\rho^{2}\sigma^{2}-\rho\sigma^{3}-\sigma\rho\sigma^{2}+\sigma
^{4})\rho\sigma\nonumber\\
&  \quad-(\rho^{4}-\rho\sigma\rho^{2}-\sigma\rho^{3}+\sigma^{2}\rho^{2}%
-\rho^{3}\sigma+\rho\sigma\rho\sigma+\sigma\rho^{2}\sigma-\sigma^{2}\rho
\sigma\nonumber\\
&  \quad-\rho^{2}\sigma\rho+\rho\sigma^{2}\rho+\sigma\rho\sigma\rho-\sigma
^{3}\rho+\rho^{2}\sigma^{2}-\rho\sigma^{3}-\sigma\rho\sigma^{2}+\sigma
^{4})\sigma\rho\nonumber\\
&  +(\rho^{4}-\rho\sigma\rho^{2}-\sigma\rho^{3}+\sigma^{2}\rho^{2}-\rho
^{3}\sigma+\rho\sigma\rho\sigma+\sigma\rho^{2}\sigma-\sigma^{2}\rho
\sigma\nonumber\\
&  \quad-\rho^{2}\sigma\rho+\rho\sigma^{2}\rho+\sigma\rho\sigma\rho-\sigma
^{3}\rho+\rho^{2}\sigma^{2}-\rho\sigma^{3}-\sigma\rho\sigma^{2}+\sigma
^{4})\sigma^{2}\\
&  =\rho^{4}\rho^{2}-\rho\sigma\rho^{2}\rho^{2}-\sigma\rho^{3}\rho^{2}%
+\sigma^{2}\rho^{2}\rho^{2}-\rho^{3}\sigma\rho^{2}+\rho\sigma\rho\sigma
\rho^{2}+\sigma\rho^{2}\sigma\rho^{2}-\sigma^{2}\rho\sigma\rho^{2}\nonumber\\
&  \quad-\rho^{2}\sigma\rho\rho^{2}+\rho\sigma^{2}\rho\rho^{2}+\sigma
\rho\sigma\rho\rho^{2}-\sigma^{3}\rho\rho^{2}+\rho^{2}\sigma^{2}\rho^{2}%
-\rho\sigma^{3}\rho^{2}-\sigma\rho\sigma^{2}\rho^{2}+\sigma^{4}\rho
^{2}\nonumber\\
&  \quad-\rho^{4}\rho\sigma+\rho\sigma\rho^{2}\rho\sigma+\sigma\rho^{3}%
\rho\sigma-\sigma^{2}\rho^{2}\rho\sigma+\rho^{3}\sigma\rho\sigma-\rho
\sigma\rho\sigma\rho\sigma-\sigma\rho^{2}\sigma\rho\sigma+\sigma^{2}\rho
\sigma\rho\sigma\nonumber\\
&  +\rho^{2}\sigma\rho\rho\sigma-\rho\sigma^{2}\rho\rho\sigma-\sigma\rho
\sigma\rho\rho\sigma+\sigma^{3}\rho\rho\sigma-\rho^{2}\sigma^{2}\rho
\sigma+\rho\sigma^{3}\rho\sigma+\sigma\rho\sigma^{2}\rho\sigma-\sigma^{4}%
\rho\sigma\nonumber\\
&  \quad-\rho^{4}\sigma\rho+\rho\sigma\rho^{2}\sigma\rho+\sigma\rho^{3}%
\sigma\rho-\sigma^{2}\rho^{2}\sigma\rho+\rho^{3}\sigma\sigma\rho-\rho
\sigma\rho\sigma\sigma\rho-\sigma\rho^{2}\sigma\sigma\rho+\sigma^{2}\rho
\sigma\sigma\rho\nonumber\\
&  +\rho^{2}\sigma\rho\sigma\rho-\rho\sigma^{2}\rho\sigma\rho-\sigma\rho
\sigma\rho\sigma\rho+\sigma^{3}\rho\sigma\rho-\rho^{2}\sigma^{2}\sigma
\rho+\rho\sigma^{3}\sigma\rho+\sigma\rho\sigma^{2}\sigma\rho-\sigma^{4}%
\sigma\rho\nonumber\\
&  +\rho^{4}\sigma^{2}-\rho\sigma\rho^{2}\sigma^{2}-\sigma\rho^{3}\sigma
^{2}+\sigma^{2}\rho^{2}\sigma^{2}-\rho^{3}\sigma\sigma^{2}+\rho\sigma
\rho\sigma\sigma^{2}+\sigma\rho^{2}\sigma\sigma^{2}-\sigma^{2}\rho\sigma
\sigma^{2}\nonumber\\
&  \quad-\rho^{2}\sigma\rho\sigma^{2}+\rho\sigma^{2}\rho\sigma^{2}+\sigma
\rho\sigma\rho\sigma^{2}-\sigma^{3}\rho\sigma^{2}+\rho^{2}\sigma^{2}\sigma
^{2}-\rho\sigma^{3}\sigma^{2}-\sigma\rho\sigma^{2}\sigma^{2}+\sigma^{4}%
\sigma^{2}%
\end{align}%
\begin{multline}
=\rho^{6}-\rho\sigma\rho^{4}-\sigma\rho^{5}+\sigma^{2}\rho^{4}-\rho^{3}%
\sigma\rho^{2}+\rho\sigma\rho\sigma\rho^{2}+\sigma\rho^{2}\sigma\rho
^{2}-\sigma^{2}\rho\sigma\rho^{2}\\
-\rho^{2}\sigma\rho^{3}+\rho\sigma^{2}\rho^{3}+\sigma\rho\sigma\rho^{3}%
-\sigma^{3}\rho^{3}+\rho^{2}\sigma^{2}\rho^{2}-\rho\sigma^{3}\rho^{2}%
-\sigma\rho\sigma^{2}\rho^{2}+\sigma^{4}\rho^{2}\\
-\rho^{5}\sigma+\rho\sigma\rho^{3}\sigma+\sigma\rho^{4}\sigma-\sigma^{2}%
\rho^{3}\sigma+\rho^{3}\sigma\rho\sigma-\rho\sigma\rho\sigma\rho\sigma
-\sigma\rho^{2}\sigma\rho\sigma+\sigma^{2}\rho\sigma\rho\sigma\\
+\rho^{2}\sigma\rho^{2}\sigma-\rho\sigma^{2}\rho^{2}\sigma-\sigma\rho
\sigma\rho^{2}\sigma+\sigma^{3}\rho^{2}\sigma-\rho^{2}\sigma^{2}\rho
\sigma+\rho\sigma^{3}\rho\sigma+\sigma\rho\sigma^{2}\rho\sigma-\sigma^{4}%
\rho\sigma\\
-\rho^{4}\sigma\rho+\rho\sigma\rho^{2}\sigma\rho+\sigma\rho^{3}\sigma
\rho-\sigma^{2}\rho^{2}\sigma\rho+\rho^{3}\sigma^{2}\rho-\rho\sigma\rho
\sigma^{2}\rho-\sigma\rho^{2}\sigma^{2}\rho+\sigma^{2}\rho\sigma^{2}\rho\\
+\rho^{2}\sigma\rho\sigma\rho-\rho\sigma^{2}\rho\sigma\rho-\sigma\rho
\sigma\rho\sigma\rho+\sigma^{3}\rho\sigma\rho-\rho^{2}\sigma^{3}\rho
+\rho\sigma^{4}\rho+\sigma\rho\sigma^{3}\rho-\sigma^{5}\rho\\
+\rho^{4}\sigma^{2}-\rho\sigma\rho^{2}\sigma^{2}-\sigma\rho^{3}\sigma
^{2}+\sigma^{2}\rho^{2}\sigma^{2}-\rho^{3}\sigma^{3}+\rho\sigma\rho\sigma
^{3}+\sigma\rho^{2}\sigma^{3}-\sigma^{2}\rho\sigma^{3}\\
-\rho^{2}\sigma\rho\sigma^{2}+\rho\sigma^{2}\rho\sigma^{2}+\sigma\rho
\sigma\rho\sigma^{2}-\sigma^{3}\rho\sigma^{2}+\rho^{2}\sigma^{4}-\rho
\sigma^{5}-\sigma\rho\sigma^{4}+\sigma^{6}.
\end{multline}

Now take the trace, giving%
\begin{multline}
\operatorname{Tr}[\rho^{6}]-\operatorname{Tr}[\rho\sigma\rho^{4}%
]-\operatorname{Tr}[\sigma\rho^{5}]+\operatorname{Tr}[\sigma^{2}\rho
^{4}]-\operatorname{Tr}[\rho^{3}\sigma\rho^{2}]+\operatorname{Tr}[\rho
\sigma\rho\sigma\rho^{2}]+\operatorname{Tr}[\sigma\rho^{2}\sigma\rho
^{2}]-\operatorname{Tr}[\sigma^{2}\rho\sigma\rho^{2}]\\
-\operatorname{Tr}[\rho^{2}\sigma\rho^{3}]+\operatorname{Tr}[\rho\sigma
^{2}\rho^{3}]+\operatorname{Tr}[\sigma\rho\sigma\rho^{3}]-\operatorname{Tr}%
[\sigma^{3}\rho^{3}]+\operatorname{Tr}[\rho^{2}\sigma^{2}\rho^{2}%
]-\operatorname{Tr}[\rho\sigma^{3}\rho^{2}]-\operatorname{Tr}[\sigma\rho
\sigma^{2}\rho^{2}]+\operatorname{Tr}[\sigma^{4}\rho^{2}]\\
-\operatorname{Tr}[\rho^{5}\sigma]+\operatorname{Tr}[\rho\sigma\rho^{3}%
\sigma]+\operatorname{Tr}[\sigma\rho^{4}\sigma]-\operatorname{Tr}[\sigma
^{2}\rho^{3}\sigma]+\operatorname{Tr}[\rho^{3}\sigma\rho\sigma
]-\operatorname{Tr}[\rho\sigma\rho\sigma\rho\sigma]-\operatorname{Tr}%
[\sigma\rho^{2}\sigma\rho\sigma]+\operatorname{Tr}[\sigma^{2}\rho\sigma
\rho\sigma]\\
+\operatorname{Tr}[\rho^{2}\sigma\rho^{2}\sigma]-\operatorname{Tr}[\rho
\sigma^{2}\rho^{2}\sigma]-\operatorname{Tr}[\sigma\rho\sigma\rho^{2}%
\sigma]+\operatorname{Tr}[\sigma^{3}\rho^{2}\sigma]-\operatorname{Tr}[\rho
^{2}\sigma^{2}\rho\sigma]+\operatorname{Tr}[\rho\sigma^{3}\rho\sigma
]+\operatorname{Tr}[\sigma\rho\sigma^{2}\rho\sigma]-\operatorname{Tr}%
[\sigma^{4}\rho\sigma]\\
-\operatorname{Tr}[\rho^{4}\sigma\rho]+\operatorname{Tr}[\rho\sigma\rho
^{2}\sigma\rho]+\operatorname{Tr}[\sigma\rho^{3}\sigma\rho]-\operatorname{Tr}%
[\sigma^{2}\rho^{2}\sigma\rho]+\operatorname{Tr}[\rho^{3}\sigma^{2}%
\rho]-\operatorname{Tr}[\rho\sigma\rho\sigma^{2}\rho]-\operatorname{Tr}%
[\sigma\rho^{2}\sigma^{2}\rho]+\operatorname{Tr}[\sigma^{2}\rho\sigma^{2}%
\rho]\\
+\operatorname{Tr}[\rho^{2}\sigma\rho\sigma\rho]-\operatorname{Tr}[\rho
\sigma^{2}\rho\sigma\rho]-\operatorname{Tr}[\sigma\rho\sigma\rho\sigma
\rho]+\operatorname{Tr}[\sigma^{3}\rho\sigma\rho]-\operatorname{Tr}[\rho
^{2}\sigma^{3}\rho]+\operatorname{Tr}[\rho\sigma^{4}\rho]+\operatorname{Tr}%
[\sigma\rho\sigma^{3}\rho]-\operatorname{Tr}[\sigma^{5}\rho]\\
+\operatorname{Tr}[\rho^{4}\sigma^{2}]-\operatorname{Tr}[\rho\sigma\rho
^{2}\sigma^{2}]-\operatorname{Tr}[\sigma\rho^{3}\sigma^{2}]+\operatorname{Tr}%
[\sigma^{2}\rho^{2}\sigma^{2}]-\operatorname{Tr}[\rho^{3}\sigma^{3}%
]+\operatorname{Tr}[\rho\sigma\rho\sigma^{3}]+\operatorname{Tr}[\sigma\rho
^{2}\sigma^{3}]-\operatorname{Tr}[\sigma^{2}\rho\sigma^{3}]\\
-\operatorname{Tr}[\rho^{2}\sigma\rho\sigma^{2}]+\operatorname{Tr}[\rho
\sigma^{2}\rho\sigma^{2}]+\operatorname{Tr}[\sigma\rho\sigma\rho\sigma
^{2}]-\operatorname{Tr}[\sigma^{3}\rho\sigma^{2}]+\operatorname{Tr}[\rho
^{2}\sigma^{4}]-\operatorname{Tr}[\rho\sigma^{5}]-\operatorname{Tr}[\sigma
\rho\sigma^{4}]+\operatorname{Tr}[\sigma^{6}]
\end{multline}

\begin{multline}
=\operatorname{Tr}[\rho^{6}]-\operatorname{Tr}[\rho^{5}\sigma
]-\operatorname{Tr}[\rho^{5}\sigma]+\operatorname{Tr}[\rho^{4}\sigma
^{2}]-\operatorname{Tr}[\rho^{5}\sigma]+\operatorname{Tr}[\rho^{3}\sigma
\rho\sigma]+\operatorname{Tr}[\rho^{2}\sigma\rho^{2}\sigma]-\operatorname{Tr}%
[\rho^{2}\sigma^{2}\rho\sigma]\\
-\operatorname{Tr}[\rho^{5}\sigma]+\operatorname{Tr}[\rho^{4}\sigma
^{2}]+\operatorname{Tr}[\rho^{3}\sigma\rho\sigma]-\operatorname{Tr}[\rho
^{3}\sigma^{3}]+\operatorname{Tr}[\rho^{4}\sigma^{2}]-\operatorname{Tr}%
[\rho^{3}\sigma^{3}]-\operatorname{Tr}[\rho^{2}\sigma\rho\sigma^{2}%
]+\operatorname{Tr}[\rho^{2}\sigma^{4}]\\
-\operatorname{Tr}[\rho^{5}\sigma]+\operatorname{Tr}[\rho^{3}\sigma\rho
\sigma]+\operatorname{Tr}[\rho^{4}\sigma^{2}]-\operatorname{Tr}[\rho^{3}%
\sigma^{3}]+\operatorname{Tr}[\rho^{3}\sigma\rho\sigma]-\operatorname{Tr}%
[\rho\sigma\rho\sigma\rho\sigma]-\operatorname{Tr}[\rho^{2}\sigma\rho
\sigma^{2}]+\operatorname{Tr}[\rho\sigma\rho\sigma^{3}]\\
+\operatorname{Tr}[\rho^{2}\sigma\rho^{2}\sigma]-\operatorname{Tr}[\rho
^{2}\sigma\rho\sigma^{2}]-\operatorname{Tr}[\rho^{2}\sigma^{2}\rho
\sigma]+\operatorname{Tr}[\rho^{2}\sigma^{4}]-\operatorname{Tr}[\rho^{2}%
\sigma^{2}\rho\sigma]+\operatorname{Tr}[\rho\sigma^{3}\rho\sigma
]+\operatorname{Tr}[\rho\sigma^{2}\rho\sigma^{2}]-\operatorname{Tr}[\rho
\sigma^{5}]\\
-\operatorname{Tr}[\rho^{5}\sigma]+\operatorname{Tr}[\rho^{2}\sigma\rho
^{2}\sigma]+\operatorname{Tr}[\rho^{3}\sigma\rho\sigma]-\operatorname{Tr}%
[\rho^{2}\sigma\rho\sigma^{2}]+\operatorname{Tr}[\rho^{4}\sigma^{2}%
]-\operatorname{Tr}[\rho^{2}\sigma\rho\sigma^{2}]-\operatorname{Tr}[\rho
^{2}\sigma^{2}\rho\sigma]+\operatorname{Tr}[\rho\sigma^{2}\rho\sigma^{2}]\\
+\operatorname{Tr}[\rho^{3}\sigma\rho\sigma]-\operatorname{Tr}[\rho^{2}%
\sigma^{2}\rho\sigma]-\operatorname{Tr}[\rho\sigma\rho\sigma\rho
\sigma]+\operatorname{Tr}[\rho\sigma\rho\sigma^{3}]-\operatorname{Tr}[\rho
^{3}\sigma^{3}]+\operatorname{Tr}[\rho^{2}\sigma^{4}]+\operatorname{Tr}%
[\rho\sigma\rho\sigma^{3}]-\operatorname{Tr}[\rho\sigma^{5}]\\
+\operatorname{Tr}[\rho^{4}\sigma^{2}]-\operatorname{Tr}[\rho^{2}\sigma
^{2}\rho\sigma]-\operatorname{Tr}[\rho^{3}\sigma^{3}]+\operatorname{Tr}%
[\rho^{2}\sigma^{4}]-\operatorname{Tr}[\rho^{3}\sigma^{3}]+\operatorname{Tr}%
[\rho\sigma\rho\sigma^{3}]+\operatorname{Tr}[\rho^{2}\sigma^{4}%
]-\operatorname{Tr}[\rho\sigma^{5}]\\
-\operatorname{Tr}[\rho^{2}\sigma\rho\sigma^{2}]+\operatorname{Tr}[\rho
\sigma^{2}\rho\sigma^{2}]+\operatorname{Tr}[\rho\sigma\rho\sigma
^{3}]-\operatorname{Tr}[\rho\sigma^{5}]+\operatorname{Tr}[\rho^{2}\sigma
^{4}]-\operatorname{Tr}[\rho\sigma^{5}]-\operatorname{Tr}[\rho\sigma
^{5}]+\operatorname{Tr}[\sigma^{6}]
\end{multline}
\bigskip%

\begin{multline}
=\operatorname{Tr}[\rho^{6}]-\operatorname{Tr}[\rho^{5}\sigma
]-\operatorname{Tr}[\rho^{5}\sigma]+\operatorname{Tr}[\rho^{4}\sigma
^{2}]-\operatorname{Tr}[\rho^{5}\sigma]+\operatorname{Tr}[\rho^{3}\sigma
\rho\sigma]+\operatorname{Tr}[\rho^{2}\sigma\rho^{2}\sigma]-\operatorname{Tr}%
[\rho^{2}\sigma^{2}\rho\sigma]\\
-\operatorname{Tr}[\rho^{5}\sigma]+\operatorname{Tr}[\rho^{4}\sigma
^{2}]+\operatorname{Tr}[\rho^{3}\sigma\rho\sigma]-\operatorname{Tr}[\rho
^{3}\sigma^{3}]+\operatorname{Tr}[\rho^{4}\sigma^{2}]-\operatorname{Tr}%
[\rho^{3}\sigma^{3}]-\operatorname{Tr}[\rho^{2}\sigma\rho\sigma^{2}%
]+\operatorname{Tr}[\rho^{2}\sigma^{4}]\\
-\operatorname{Tr}[\rho^{5}\sigma]+\operatorname{Tr}[\rho^{3}\sigma\rho
\sigma]+\operatorname{Tr}[\rho^{4}\sigma^{2}]-\operatorname{Tr}[\rho^{3}%
\sigma^{3}]+\operatorname{Tr}[\rho^{3}\sigma\rho\sigma]-\operatorname{Tr}%
[\rho\sigma\rho\sigma\rho\sigma]-\operatorname{Tr}[\rho^{2}\sigma\rho
\sigma^{2}]+\operatorname{Tr}[\rho\sigma\rho\sigma^{3}]\\
+\operatorname{Tr}[\rho^{2}\sigma\rho^{2}\sigma]-\operatorname{Tr}[\rho
^{2}\sigma\rho\sigma^{2}]-\operatorname{Tr}[\rho^{2}\sigma^{2}\rho
\sigma]+\operatorname{Tr}[\rho^{2}\sigma^{4}]-\operatorname{Tr}[\rho^{2}%
\sigma^{2}\rho\sigma]+\operatorname{Tr}[\rho\sigma^{3}\rho\sigma
]+\operatorname{Tr}[\rho\sigma^{2}\rho\sigma^{2}]-\operatorname{Tr}[\rho
\sigma^{5}]\\
-\operatorname{Tr}[\rho^{5}\sigma]+\operatorname{Tr}[\rho^{2}\sigma\rho
^{2}\sigma]+\operatorname{Tr}[\rho^{3}\sigma\rho\sigma]-\operatorname{Tr}%
[\rho^{2}\sigma\rho\sigma^{2}]+\operatorname{Tr}[\rho^{4}\sigma^{2}%
]-\operatorname{Tr}[\rho^{2}\sigma\rho\sigma^{2}]-\operatorname{Tr}[\rho
^{2}\sigma^{2}\rho\sigma]+\operatorname{Tr}[\rho\sigma^{2}\rho\sigma^{2}]\\
+\operatorname{Tr}[\rho^{3}\sigma\rho\sigma]-\operatorname{Tr}[\rho^{2}%
\sigma^{2}\rho\sigma]-\operatorname{Tr}[\rho\sigma\rho\sigma\rho
\sigma]+\operatorname{Tr}[\rho\sigma\rho\sigma^{3}]-\operatorname{Tr}[\rho
^{3}\sigma^{3}]+\operatorname{Tr}[\rho^{2}\sigma^{4}]+\operatorname{Tr}%
[\rho\sigma\rho\sigma^{3}]-\operatorname{Tr}[\rho\sigma^{5}]\\
+\operatorname{Tr}[\rho^{4}\sigma^{2}]-\operatorname{Tr}[\rho^{2}\sigma
^{2}\rho\sigma]-\operatorname{Tr}[\rho^{3}\sigma^{3}]+\operatorname{Tr}%
[\rho^{2}\sigma^{4}]-\operatorname{Tr}[\rho^{3}\sigma^{3}]+\operatorname{Tr}%
[\rho\sigma\rho\sigma^{3}]+\operatorname{Tr}[\rho^{2}\sigma^{4}%
]-\operatorname{Tr}[\rho\sigma^{5}]\\
-\operatorname{Tr}[\rho^{2}\sigma\rho\sigma^{2}]+\operatorname{Tr}[\rho
\sigma^{2}\rho\sigma^{2}]+\operatorname{Tr}[\rho\sigma\rho\sigma
^{3}]-\operatorname{Tr}[\rho\sigma^{5}]+\operatorname{Tr}[\rho^{2}\sigma
^{4}]-\operatorname{Tr}[\rho\sigma^{5}]-\operatorname{Tr}[\rho\sigma
^{5}]+\operatorname{Tr}[\sigma^{6}]
\end{multline}

\begin{align}
& \!\!\!\! =\operatorname{Tr}[\rho^{6}]\nonumber\\
&  -\operatorname{Tr}[\rho^{5}\sigma]-\operatorname{Tr}[\rho^{5}%
\sigma]-\operatorname{Tr}[\rho^{5}\sigma]-\operatorname{Tr}[\rho^{5}%
\sigma]-\operatorname{Tr}[\rho^{5}\sigma]-\operatorname{Tr}[\rho^{5}%
\sigma]\nonumber\\
&  +\operatorname{Tr}[\rho^{4}\sigma^{2}]+\operatorname{Tr}[\rho^{4}\sigma
^{2}]+\operatorname{Tr}[\rho^{4}\sigma^{2}]+\operatorname{Tr}[\rho^{4}%
\sigma^{2}]+\operatorname{Tr}[\rho^{4}\sigma^{2}]+\operatorname{Tr}[\rho
^{4}\sigma^{2}]\nonumber\\
&  +\operatorname{Tr}[\rho^{3}\sigma\rho\sigma]+\operatorname{Tr}[\rho
^{3}\sigma\rho\sigma]+\operatorname{Tr}[\rho^{3}\sigma\rho\sigma
]+\operatorname{Tr}[\rho^{3}\sigma\rho\sigma]+\operatorname{Tr}[\rho^{3}%
\sigma\rho\sigma]+\operatorname{Tr}[\rho^{3}\sigma\rho\sigma]\nonumber\\
&  +\operatorname{Tr}[\rho^{2}\sigma\rho^{2}\sigma]+\operatorname{Tr}[\rho
^{2}\sigma\rho^{2}\sigma]+\operatorname{Tr}[\rho^{2}\sigma\rho^{2}%
\sigma]\nonumber\\
&  -\operatorname{Tr}[\rho^{2}\sigma^{2}\rho\sigma]-\operatorname{Tr}[\rho
^{2}\sigma^{2}\rho\sigma]-\operatorname{Tr}[\rho^{2}\sigma^{2}\rho
\sigma]-\operatorname{Tr}[\rho^{2}\sigma^{2}\rho\sigma]-\operatorname{Tr}%
[\rho^{2}\sigma^{2}\rho\sigma]-\operatorname{Tr}[\rho^{2}\sigma^{2}\rho
\sigma]\nonumber\\
&  -\operatorname{Tr}[\rho^{3}\sigma^{3}]-\operatorname{Tr}[\rho^{3}\sigma
^{3}]-\operatorname{Tr}[\rho^{3}\sigma^{3}]-\operatorname{Tr}[\rho^{3}%
\sigma^{3}]-\operatorname{Tr}[\rho^{3}\sigma^{3}]-\operatorname{Tr}[\rho
^{3}\sigma^{3}]\nonumber\\
&  -\operatorname{Tr}[\rho^{2}\sigma\rho\sigma^{2}]-\operatorname{Tr}[\rho
^{2}\sigma\rho\sigma^{2}]-\operatorname{Tr}[\rho^{2}\sigma\rho\sigma
^{2}]\nonumber\\
&  -\operatorname{Tr}[\rho\sigma\rho\sigma\rho\sigma]-\operatorname{Tr}%
[\rho\sigma\rho\sigma\rho\sigma]\nonumber\\
&  +\operatorname{Tr}[\rho\sigma\rho\sigma^{3}]+\operatorname{Tr}[\rho
\sigma\rho\sigma^{3}]+\operatorname{Tr}[\rho\sigma\rho\sigma^{3}%
]+\operatorname{Tr}[\rho\sigma\rho\sigma^{3}]+\operatorname{Tr}[\rho\sigma
\rho\sigma^{3}]+\operatorname{Tr}[\rho\sigma\rho\sigma^{3}]\nonumber\\
&  -\operatorname{Tr}[\rho^{2}\sigma\rho\sigma^{2}]-\operatorname{Tr}[\rho
^{2}\sigma\rho\sigma^{2}]-\operatorname{Tr}[\rho^{2}\sigma\rho\sigma
^{2}]\nonumber\\
&  +\operatorname{Tr}[\rho\sigma^{2}\rho\sigma^{2}]+\operatorname{Tr}%
[\rho\sigma^{2}\rho\sigma^{2}]+\operatorname{Tr}[\rho\sigma^{2}\rho\sigma
^{2}]\nonumber\\
&  +\operatorname{Tr}[\rho^{2}\sigma^{4}]+\operatorname{Tr}[\rho^{2}\sigma
^{4}]+\operatorname{Tr}[\rho^{2}\sigma^{4}]+\operatorname{Tr}[\rho^{2}%
\sigma^{4}]+\operatorname{Tr}[\rho^{2}\sigma^{4}]+\operatorname{Tr}[\rho
^{2}\sigma^{4}]\nonumber\\
&  -\operatorname{Tr}[\rho\sigma^{5}]-\operatorname{Tr}[\rho\sigma
^{5}]-\operatorname{Tr}[\rho\sigma^{5}]-\operatorname{Tr}[\rho\sigma
^{5}]-\operatorname{Tr}[\rho\sigma^{5}]-\operatorname{Tr}[\rho\sigma
^{5}]\nonumber\\
&  +\operatorname{Tr}[\sigma^{6}].
\end{align}
This reduces to%
\begin{multline}
=\operatorname{Tr}[\rho^{6}]-6\operatorname{Tr}[\rho^{5}\sigma
]+6\operatorname{Tr}[\rho^{4}\sigma^{2}]+6\operatorname{Tr}[\rho^{3}\sigma
\rho\sigma]+3\operatorname{Tr}[\rho^{2}\sigma\rho^{2}\sigma
]-6\operatorname{Tr}[\rho^{2}\sigma^{2}\rho\sigma]\\
-6\operatorname{Tr}[\rho^{3}\sigma^{3}]-3\operatorname{Tr}[\rho^{2}\sigma
\rho\sigma^{2}]-2\operatorname{Tr}[\rho\sigma\rho\sigma\rho\sigma
]+6\operatorname{Tr}[\rho\sigma\rho\sigma^{3}]-3\operatorname{Tr}[\rho
^{2}\sigma\rho\sigma^{2}]\\
+3\operatorname{Tr}[\rho\sigma^{2}\rho\sigma^{2}]+6\operatorname{Tr}[\rho
^{2}\sigma^{4}]-6\operatorname{Tr}[\rho\sigma^{5}]+\operatorname{Tr}%
[\sigma^{6}],
\end{multline}
thus implying the claimed formula in \eqref{eq:schatten-6}.

\end{document}